\documentclass[12pt,a4paper]{article}
\usepackage{graphicx}
\title{On electronic shells surrounding charged insulated metallic 
clusters
}

\author{K. Dietrich and M. Garny \\
{\it Physikdepartment of the Technische Universit\"at M\"unchen}\\
{\it D-85747 Garching (FRG)} \\
K. Pomorski \\
{\it Katedra Fizyki Teoretycznej, Uniwersytet Marii Curie-Sk\l odowskiej}
\\
{\it PL-20031 Lublin (Poland)}}

\begin{document}

\maketitle

\begin{abstract}
We determine the wavefunctions of electrons bound to a positively charged
mesoscopic metallic cluster covered by an insulating surface layer. The
radius of the metal core and the thickness of the insulating surface layer
are of the order of a couple of {\aa}ngstr\"om. We study in particular the
electromagnetic decay of externally located electrons into unoccupied
internally located states which exhibits a resonance behaviour. This
resonance structure has the consequence that the lifetime of the
``mesoscopic atoms'' may vary by up to 6 orders of magnitude depending on 
the values of the parameters (from sec to years).

\end{abstract}

\section{Introduction}
\label{intro}

Clusters consisting of some 10$^2$ to 10$^5$ atoms, which can be
produced for instance in the course of the condensation of
vapour, have attracted great interest during the last 20 years.
These ''mesoscopic'' clusters were found to exhibit classical as
well as quantum mechanical features. In the special case of
metallic clusters \cite{Kni84}, the conduction electrons, which move in the
average potential produced by the positively charged ions,
exhibit shell effects. As a consequence, the binding energy of
the cluster becomes a function of the number of free electrons
(and thus also of the number of ions). In turn, this implies
measurable variations of the formation probability of a cluster
as a function of its mass in a thermal vapour-gas equilibrium.

Other interesting types of mesoscopic systems are the so-called
''fullerenes'' which represent a crystal of atoms covering a
closed, usually spherical surface \cite{Fra97}. A number of review
articles exist on this exciting new field of physics (see for
instance Ref.~\cite{Bra93} and Ref.~\cite{Hee93} and books mentioned at the 
end of Ref.~\cite{Kni84} and Ref.~\cite{Fra97}).

In the present paper, we consider a mesoscopic system which
consists of a spherical metallic core of radius $R_2$ carrying
$Z$ positive charges and surrounded by an insulating layer of
external radius $R_1$ (see Fig.~1). We determine analytically
the bound states of a single electron in the potential $V(r)$
which is produced by the positive surface charge density of the
metallic core and the insulating dielectric surface layer. As in
the case of an ordinary atom, these single electron
wave functions describe also approximately the case that several
electrons are bound in this potential. For $Z$ bound electrons,
the system has total charge 0. Because of the obvious similarity
of this system with an atom or ion, we refer to it as a
''mesoscopic atom'' or ''mesoscopic ion''.

As one can surmise, there are two different types of bound
electronic states: electrons localized in the metallic core
inside of the insulating layer (region II, Fig.~1) and
electrons localized in the vacuum outside of the insulating
layer (region 0, Fig.~1). The bound states localized outside
are reminiscent of the bound states of an electron in the
Coulomb potential of the atomic nucleus. We are mainly
interested in these externally localized electronic states
(''class 1-states'') and their decay into lower-lying bound states
predominantly localized within the metallic core (''class 2-states'').

The paper is organized as follows:

In Sec.~2, we formulate the single-particle Hamiltonian and
determine the eigenfunctions and eigenenergies of bound
electrons. In doing so, we take into account the polarization of the
insulating surface layer which is produced by the uniform
positive charge distribution on the surface $S_2$ of the metal,
but we neglect the additional polarization effects produced by
the externally localized electrons.

In Sec.~3 we determine these additional polarization effects
within the classical Maxwellian electrodynamics.  They lead to
an additional potential acting on the electrons which is of
importance close to the surface $S_1$ of the insulator. So far,
we have not yet studied the effect of this additional potential
on the externally localized electrons.

The Sec.~4 is dedicated to a presentation of the
results. In Subsec.~4.1, we describe the procedure which we used
in order to obtain the eigenvalues and eigenfunctions and we
present typical examples of spectra and wavefunctions. In Subsec.~4.2, 
we discuss the electromagnetic decay of a predominantly
externally localized bound electron into a lower-lying internally
localized state. We show that the transition rate and the
corresponding life-time may vary by several orders of magnitude
depending on the parameters of the system.

In general, the transition is hindered by the small overlap between externally
and internally localized states (''class~1-states'' and ''class~2-states'') if
the insulating surface layer is sufficiently thick. In particular, we discuss
the special role in this decay of ''class 3-states'' which exhibit
non-negligible amplitudes both in the external region 0 and the internal region
II.  

Finally, in Sec. 5, we summarize the results and discuss some
related open questions.

\begin{figure}
\label{fig1}
\begin{center}
\includegraphics[width=7cm]{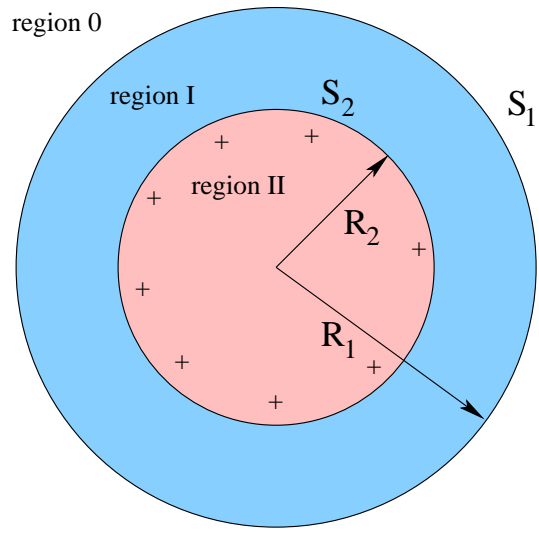}
\caption{Schematic view of the charged metallic cluster (radius
$R_2$, surface $S_2$) covered by an insulating layer (outer
radius $R_1$, surface $S_1$). States of bound electrons
predominantly localized in the surrounding vacuum ($R > R_1$,
region 0) are called ''class 1-states'' in distinction from states
localized in the metallic core (''class 2-states'').}
\end{center}
\end{figure}

\section{Theory}
\label{sec:2}

In the single-particle Hamiltonian for an electron bound to a
positively charged and insulated metallic core ($M$ = mass of
the electron)
\begin{equation}
 \widehat H = \widehat T + \widehat V = -{\hbar^2\over 2M} \Delta + V(r) \,\,,
\label{eq2.1}
\end{equation}
the potential $V(r)$ contains the interaction of the electron
with the positive surplus charge $Ze_0$ ($e_0$ = elementary
charge) and with the ionic background (''jellium'') of the metal. Furthermore,
it contains the effects of the insulating surface layer. The radius
$R_2$ of the metal and the thickness ($R_1 - R_2$) of the
surface layer are assumed to be of the order of 5 to 20 \AA (1
\AA = 10$^{-10}$~m) and the number $Z$ of positive charges of the metal 
is considered to be 1 to 5. For these dimensions of the system 
we may expect that, on the one hand, the dynamics of the electron is fully 
quantum-mechanical and, on the other hand, we may use macroscopic concepts 
in formulating the potential $V(r)$, at least as a first approximation. 
So we assume that the positive surplus charge is uniformly distributed 
over the surface $S_2$ of the metal
\begin{equation}
 \rho_2(r) = \rho_{S_2} \cdot \delta (r - R_2) = {Ze_0\over 4\pi R^2_2} 
 \delta(r - R_2)\,.
\label{eq2.2}
\end{equation}
The Coulomb potential produced by this simple charge
distribution acting upon an electron (charge $-e_0$) is given by
\begin{equation}
 -e_0\int d^3 r' {\rho_2(r')\over |\vec r - \vec r'|} 
   = -{Ze^2_0\over R_2}\,
 \theta_0(R_2 - r) - {Ze^2_0\over r} \theta_0(r - R_2) \,\,,
\label{eq2.3}
\end{equation}
where $\theta_0(x)$ is the Heaviside function
$$ 
 \theta_0(x) = \left\{\begin{array}{lll}
1& {\rm for} & x > 0 \\ 0 &{\rm for} & x < 0\end{array}\right. \,\,.
$$ 

Inside of the metal, the conduction electrons feel not only the constant
potential produced by the positive surface charge but also the attractive
interaction with the ionic background which we represent by a constant negative
potential $-V^0_{\rm II}$ $(V^0_{\rm II} > 0$).

On the other hand, the average energy gap above the Fermi level in the
insulator may act upon the electron like a repulsive barrier of a height 
$B_0 > 0$. 

Last not least, there are polarization effects produced by the positive surplus
charge and also by the charge distribution $\rho_e(\vec r)$ of the externally
(i.e. predominantly in region 0) localized electrons. The polarization effects
are investigated in Sec.~3 on the basis of Maxwell's equations for the electric
field $\vec E$ and for the displacement field $\vec D$ in the presence of
homogeneous dielectric media. In the case that we only take into account the
polarization effects produced by the positive surplus charge (2), the following
modifications of the Coulomb potential (3) are to be considered: Within region
I, i.e.  within the insulating layer, the Coulomb potential $-Ze^2_0/ r$
is to be replaced by
$$ 
 -{Ze^2_0\over r} \rightarrow - {Ze^2_0\over r \cdot \varepsilon_{\rm I}} - 
{Ze^2_0 \cdot (\varepsilon_{\rm I} - 1)\over R_1 \cdot \varepsilon_{\rm I}}
$$
and within region II, i.e. within the metallic core, the constant potential
$-{Ze^2_0/R_2}$ is to be supplanted by
$$
  -{Ze^2_0\over R_2}\, \rightarrow \, - {Ze^2_0\over R_2\varepsilon_{\rm I}} - 
  {Ze^2_0 \cdot (\varepsilon_{\rm I} - 1)\over R_1\varepsilon_{\rm I}} =
 -{Ze^2_0\over R_2} + Ze^2_0 \left({1\over R_2} - {1\over R_1}\right)
  \left({\varepsilon_{\rm I} - 1\over\varepsilon_{\rm I}}\right)\,.
$$
Within region 0, i.e. in the vacuum surrounding the insulating layer, the
value $-{Ze^2_0/r}$ of the Coulomb potential (3) is unchanged.

One should note that the potential in region I assumes the value 
$-{Ze^2_0/R_1}$ at the outer surface $r = R_1$ of the insulating layer.

Let us denote the sum of the constant parts of the potentials in region I
and II by
\begin{equation}
 B: = B_0 - {Ze^2_0\over \varepsilon_{\rm I}R_1} (\varepsilon_{\rm I} - 1)
\label{eq2.4}
\end{equation}
and
\begin{equation}
 V_{\rm II}: = V^0_{\rm II} - Ze^2_0  \left({1\over R_2} - {1\over R_1}\right)
\left({\varepsilon_{\rm I} - 1\over \varepsilon_{\rm I}}\right)\,.
\label{eq2.5}
\end{equation}
If $Z$ is equal to 1 or 2 and the dielectric constant $\varepsilon_{\rm I}$ 
exceeds 1 only slightly, and if the radii are in the range of values treated
in Sec.~4, the 2$^{\rm nd}$ terms on the right-hand sides of Eqs. 
(\ref{eq2.4}) and (\ref{eq2.5}) are much smaller than the 1$^{\rm st}$ terms. 
So far, it is only these cases which we treated numerically.

Altogether, the potential $V(r)$ felt by the electron has the form
\begin{equation}
 V(r) = -\left[V_{\rm II} + {Ze^2_0\over R_2}\right] \theta_0(R_2 - r)
 + \left[B - {Ze^2_0\over \varepsilon_{\rm I}r}\right] \theta_0(R_1 - r) 
\theta_0(r - R_2) - {Ze^2_0\over r} \theta_0(r - R_1)\,\,.
\label{eq2.5.1}
\end{equation}
The constants $V_{\rm II}$, $\varepsilon_{\rm I}$ and $B_0$ are
empirical parameters. The potential $V(r)$ is shown in Fig.~2
for a specific choice of the parameters.

\begin{figure}
\label{fig2}
\begin{center}
\includegraphics[width=7cm]{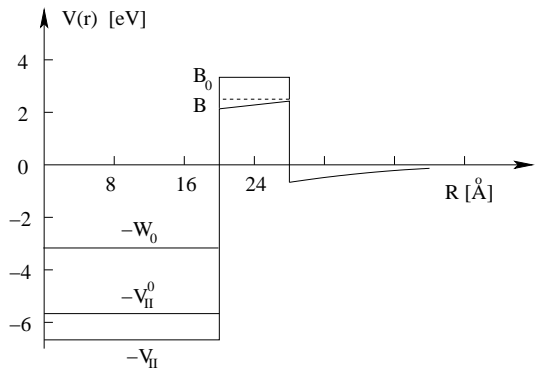}
\caption{Semi-quantitative picture of the potential $V(r)$ for
the following choice of parameters: $R_2=20$~\AA; $R_1=30$~\AA; $Z=1$; 
$W_0=3.3$~eV; $V_{\rm II}=6.3$~eV; $B=3$~eV; $\varepsilon_{\rm I}=3$ ;
$Ze^2_0/(\varepsilon_{\rm I}\cdot R_2)=0.24~{\rm eV};  
Ze^2_0/(\varepsilon_{\rm I}\cdot R_1)=0.16~{\rm eV};  
Ze^2_0/R_1=0.48~{\rm eV}\,.$}
\end{center}
\end{figure}

The eigenfunctions $\psi(r,\vartheta,\varphi)$ of the Schr\"odinger equation
\begin{equation}
 -{\hbar^2\over 2M}\Delta\psi + V(r) \psi = E\psi
\label{eq2.6}
\end{equation}
separate
\begin{equation}
 \psi(r,\vartheta,\varphi) = R(r) Y_{lm}(\vartheta,\varphi) = {u(r) \over r}
 Y_{lm}(\vartheta,\varphi)
\label{eq2.7}
\end{equation}
and the radial functions $R(r)$ and $u(r)$ have to satisfy the
differential equations
\begin{equation}
 -{\hbar^2\over 2M} \left({d^2\over dr^2} + {2\over r} {d\over dr} - 
 {l(l + 1)\over r^2}\right) R + V(r) R = ER  \,\,,
\label{eq2.8}
\end{equation}
%
\begin{equation}
 -{\hbar^2\over 2M} \left({d^2\over dr^2} - {l(l + 1)\over r^2}\right) u
   + V(r)u = Eu \,\,,
\label{eq2.9}
\end{equation}
respectively.

\underline{For $r < R_2$ (region II)}, the potential $V(r)$ is constant and
thus the Eq.~(\ref{eq2.8}) represents the differential equation for
''spherical Bessel functions''. Of the 2 linearly independent
solutions, only the regular one, $j_l$, qualifies as a physical
solution. Up to a normalization factor which will be determined
later on, the radial function $R(r)$ in region II is simply given
by
\begin{equation}
 R_{\rm II}(r) = j_l(\kappa_{\rm II}r)\,\,,
\label{eq2.10}
\end{equation}
where $\kappa_{\rm II}$ is defined by
\begin{equation}
 \kappa_{\rm II} = \sqrt{{2M\over\hbar^2} \left({Ze^2_0\over R_2} + 
 V_{\rm II} - |E|\right)}
\label{eq2.11}
\end{equation}
and $j_l(x)$ is the spherical Bessel function of the 1$^{\rm st}$ 
kind (see Chapt. 9 of Ref.~\cite{Abram}). As bound states have an
energy between the well depth $-\left(V_{\rm II} + Ze^2_0/ R_2 \right)$ and 0, 
the parameter $\kappa_{\rm II}$ is assured to be real.

\underline{For $R_2 < r < R_1$ (region I)} let us postpone the
treatment of the general case and first turn to the simpler
situation when the $r$-dependent term $Ze^2_0/(\varepsilon_{\rm I}r)$ is 
negligible in comparison to the constant $B$ in the entire region I. This is 
so, if the inequality
\begin{equation}
 {Ze^2_0\over\varepsilon_{\rm I}R_2} << B
\label{eq2.12}
\end{equation}
holds. For the case shown in Fig.~2, the condition is
fulfilled. Generally speaking, the barrier heights $B_0$ range
from 2 to 5 eV, and the dielectric constant $\varepsilon_{\rm
I}$ may have values between slightly more than 1 up to more
than 100 (Examples: $\varepsilon_{\rm I}$ (caoutchouc) = 4;
$\varepsilon_{\rm I}$ (Ti0$_2$-crystal) = 115).

Let us now investigate the case that the potential (\ref{eq2.5.1}) can be
approximated by
\begin{equation}
 V(r) \approx - \left[V_{\rm II} + {Ze^2_0\over R_2}\right] \theta_0(R_2 - r) 
 +B\theta_0 (R_1 - r) \theta_0(r - R_2) - {Ze^2_0\over r}\theta_0(r - R_1)\,\,.
\label{eq2.13}
\end{equation}
The radial Schr\"odinger equation (\ref{eq2.8}) in region I then assumes
again the form of Bessel's differential equation. But as the
energy $E$ is lower than the barrier height $B$, the solution in
region I is given by a linear combination of two independent
''modified spherical Bessel functions'' (see Chapt. 10.2 of Ref.
\cite{Abram}).

We choose the pair $\left\{(-i)^l j_l(iy), -{\pi\over 2} i^l
h^{(1)}_l(iy)\right\}$, where $j_l(iy)$ and $h^{(1)}_l(iy)$ are the
spherical Bessel-function and spherical Hankel-function of the
1$^{\rm st}$ kind, resp., with imaginary arguments. The
dimensionless real variable $y$ is related to the radial
coordinate $r$ by
\begin{equation}
 y = \kappa_{\rm I}r \,\,,
\label{eq2.14}
\end{equation}
%
\begin{equation}
 \kappa_{\rm I} = \sqrt{{2M\over\hbar^2}(B + |E|)}\,\,.
\label{eq2.15}
\end{equation}
The two factors $(-i)^l$ and $-{\pi\over 2}i^l$ are chosen so
that the pair of solutions are simply related to the modified
Bessel function of integer order $I_{l+{1\over 2}}(y)$ and
$K_{l+{1\over 2}}(y)$ for which a number of rapidly converging
series expansions exist (see Ref.~\cite{Abram}, Chapt. 10.2)
\begin{equation}
 \{(-i)^l j_l(iy), -{\pi\over 2}i^l h^{(1)}_l(iy)\}= \left\{\sqrt{{\pi\over 2y}}
 I_{l+{1\over 2}}(y), \sqrt{{\pi\over 2y}}K_{l+{1\over 2}}(y)\right\}\,\,.
\label{eq2.16}
\end{equation}
The radial function $R(r)$ in region I thus has the general form
\begin{equation}
 R_{\rm I}(r) = A_{\rm I}\sqrt{{\pi\over 2\kappa_{\rm I}r}}I_{l+{1\over 2}} 
 (\kappa_{\rm I}r) + B_{\rm I}\sqrt{{\pi\over 2\kappa_{\rm I}r}} 
  K_{l+{1\over 2}} (\kappa_{\rm I}r)\,\,.
\label{eq2.17}
\end{equation}

\medskip
\underline{For $r > R_1$ (region 0)}, it is more convenient to deal with the 
differential Eq.~(\ref{eq2.9}) for $u(r)$ which has the form
\begin{equation}
 -{\hbar^2\over 2M} u''(r) + {\hbar^2l(l+1)\over 2Mr^2} u 
 - {Ze^2_0\over r} u = Eu\,\,.
\label{eq2.18}
\end{equation}
We have to find a normalizable solution of (\ref{eq2.18}). It turns out that 
the only normalizable solution of (\ref{eq2.18}) has the form
\begin{equation}
 u(r) = C_0 z^{l+1}_0 U(a_0,b_0,z_0) e^{-{z_0\over 2}}\,\,,
\label{eq2.19}
\end{equation}
where the dimensionless variable $z_0$ is related to $r$ by
\begin{equation}
 z_0 = 2\lambda r
\label{eq2.20}
\end{equation}
with
\begin{equation}
 \lambda = \sqrt{{2M|E|\over\hbar^2}}
\label{eq2.21}
\end{equation}
and where $w = U(a_0,b_0, z_0)$ is a particular solution Kummer's 
differential equation
\begin{equation}
 z_0 {d^2w\over dz^2_0} + (b_0 - z_0) {dw\over dz_0} - a_0w = 0\,\,.
\label{eq2.22}
\end{equation}
The parameters $a_0$ and $b_0$ in (\ref{eq2.22}) are related to the parameters in 
Eq.~(\ref{eq2.18}) by 
\begin{equation}
 a_0 = {2(l + 1) - d_0\over 2} 
\label{eq2.23.1}
\end{equation}
%
\begin{equation}
 b_0 = 2 (l + 1) 
\label{eq2.23.2}
\end{equation}
%
\begin{equation}
 d_0 = {Ze^2_0\over \hbar} \sqrt{{2M\over |E|}}
\label{eq2.23.3}
\end{equation}
and the solution $U(a_0,b_0,z_0)$ of (\ref{eq2.22}) is defined as follows:\\
We write $b_0$ in the form
\begin{equation}
 b_0 = 1 + n \,\,,
\label{eq2.24.1}
\end{equation}
which implies
\begin{equation}
 n = 2l + 1 \,\,.
\label{eq2.24.2}
\end{equation}
Then the function $U(a_0, 1 + n,z_0)$, where $n = 1,\ldots$ reads 
(see Ref.~\cite{Abram}, Eq.~(13.1.6) or Ref.~\cite{Erdel}, Eq.~(13) of 
Chapt.~6.7.1)
\begin{eqnarray}
 U(a_0,n+1,z_0) &=& {(-1)^{n-1}\over n!\Gamma(a_0 - n)} \cdot 
 \left\{M(a_0, n + 1, z_0) \ln z_0 + \right. \nonumber \\
&+& \left. \sum^\infty_{\nu = 0} {(a_0)_\nu z^\nu_0\over (n+1)_\nu \nu!} 
 [\psi(a_0 + \nu) - \psi (1 + \nu) - \psi (1 + n + \nu)]\right\}\nonumber \\
&+& {(n-1)!\over \Gamma(a_0)} \sum^{n-1}_{\nu=0} {(a_0-n)_\nu\over 
 (1 - n)_\nu} \cdot {z^{\nu - n}_0\over \nu!}
\label{eq2.25}
\end{eqnarray}
In (\ref{eq2.25}), the function $\psi(x)$ is the logarithmic derivative 
of the $\Gamma$-function
\begin{equation}
 \psi(x): = {d\Gamma(x)\over dx} / \Gamma(x) = {d\ln\Gamma(x) \over dx}
\label{eq2.26}
\end{equation}
and $(c)_\nu$ is Pochhammer's symbol
\begin{equation}
 (c)_\nu: = c(c + 1) \ldots (c + \nu - 1) = {\Gamma(c + \nu)\over \Gamma(c)}\,\,.
\label{eq2.27}
\end{equation}
Furthermore, $M(a_0,b_0,z_0)$ is the confluent hypergeometric function
\begin{equation}
M(a_0,b_0,z_0) = \sum^\infty_{\nu=0} {\Gamma(a_0 + \nu)\Gamma(b_0)\over
\Gamma(a_0) \Gamma(b_0 + \nu)} \cdot {z^\nu_0\over \nu!}\,\,.
\label{eq2.28}
\end{equation}
We note in passing that the complicated form of the solution
(\ref{eq2.25}) is due to the fact that the parameter $b_0$ in $U(a_0,b_0,z_0)$ 
is an integer. For non-integer $b_0$, the function $U(a_0,b_0,z_0)$ depends
in a simple way on two confluent hypergeometric functions (see
Ref.~\cite{Abram}, Eq.~(13.1.3)). This form of $U(a_0,b_0,z_0)$ becomes
singular if $b_0$ tends to an integer value. The limiting process
which leads to the form (\ref{eq2.25}) is discussed in Ref.~\cite{Erdel}.

For any value of $b_0$, the asymptotic behaviour of $U$ for
$\Re(z_0)\rightarrow\infty$ is given by
\begin{equation}
 U(a_0,b_0,z_0) \rightarrow {1\over z^{a_0}_0} \left[ 1 + 0\left({1\over |z_0|}
 \right)\right]
\label{eq2.29}
\end{equation}
(see Ref.~\cite{Abram}, Eq.~(13.1.8)).  From (\ref{eq2.29}) and (\ref{eq2.19}) it is
clear that $u(r)$ is a normalizable function irrespectively of
the value of the energy $E$. However, it turns out that the
continuity conditions can only be met for certain discrete
values of $E$. The radial function $R_0(r)$ in region 0 is thus
defined by
\begin{equation}
 R_0(r) = {u(r)\over 2\lambda r} 
        = C_0(2\lambda r)^l U(a_0,2l+2,2\lambda r) e^{-\lambda r}\,\,.
\label{eq2.30}
\end{equation}
Before we proceed to the explicit formulation of the continuity
conditions, we return to the form of the solution in region I in
the general case that the $r$--dependent term
$Ze^2_0/(\varepsilon_{\rm I}r)$ is \underline{not}
neglected. Then the differential equation for $u(r)$ in region I
has the form
\begin{equation}
 -{\hbar^2\over 2M} \left({d^2\over dr^2} - {l(l+1)\over r^2}\right) u + 
\left(B - {Ze^2_0\over \varepsilon_{\rm I}r}\right)u = Eu \,\,.
\label{eq2.31}
\end{equation}
Eq.~(\ref{eq2.31}) resembles Eq.~(\ref{eq2.18}). In fact, with the replacements
\begin{equation}
 Z \rightarrow {Z\over\varepsilon_{\rm I}} \,\,,
\label{eq2.32.1}
\end{equation}
%
\begin{equation}
 E \rightarrow E - B  \,\,,
\label{eq2.32.2}
\end{equation}
Eq.~(\ref{eq2.18}) becomes Eq. (\ref{eq2.31}).\\
With the ansatz
\begin{equation}
 u(r) = z^{l+1}_{\rm I} w(z_{\rm I}) e^{-{Z_{\rm I}\over 2}}
\label{eq2.33}
\end{equation}
we obtain again Kummer's differential equation for $w(z)$
\begin{equation}
 z_{\rm I} {d^2w\over dz^2_{\rm I}} + (b_0 - z_{\rm I}) 
 {dw\over dz_{\rm I}} - a_{\rm I}w = 0\,,
\label{eq2.34}
\end{equation}
where $b_0$ is again given by (\ref{eq2.23.2}) or (\ref{eq2.24.1}), but $a_{\rm
I}$ is obtained from $a_0$ by the replacements (\ref{eq2.32.1}),
(\ref{eq2.32.2}):
\begin{equation}
a_{\rm I} = {2(l+1) - d_{\rm I}\over 2} \,\,,
\label{eq2.35.1}
\end{equation}
%
\begin{equation}
d_{\rm I} = {Ze^2_0\over \varepsilon_{\rm I}\hbar}\sqrt{{2M\over |E - B|}}
\label{eq2.35.2}
\end{equation}
and the dimensionless variable $z_{\rm I}$ is related to $r$ by
\begin{equation}
 z_{\rm I} = 2\cdot \sqrt{{2M \cdot |E - B|\over\hbar^2}} r 
 = 2\kappa_{\rm I}r \,\,.
\label{eq2.35.3}
\end{equation}
The parameter $\kappa_{\rm I}$ defined in (\ref{eq2.15}) agrees with
$\sqrt{2M/\hbar^2 \cdot |E-B|}$ because $E < 0$ and $B >
0$. The general solution $w$ of Eq.~(\ref{eq2.34}) can be written as a
linear combination of the function $U(a_{\rm I}, 2l+2, z_{\rm I}$)
and, as linearly independent 2$^{\rm nd}$ solution, the
confluent hypergeometric function $M(a_{\rm I}, 2l+2, z_{\rm I})$:
\begin{equation}
 w(z_{\rm I}) = \tilde A_{\rm I} M(a_{\rm I}, 2l+2, z_{\rm I}) + 
 \tilde B_{\rm I} U(a_{\rm I}, 2l+2, z_{\rm I})\,\,.
\label{eq2.36}
\end{equation}
We note that we had to suppress the solution $M(a_0, 2l+2, z_0)$ in
region 0 because it would lead to a non-normalizable solution.

The radial solution $R(r)$ in region I thus assumes the form
\begin{equation}
 R_{\rm I}(r) = {u(r)\over 2\kappa_{\rm I}r} \,\,,
\label{eq2.37}
\end{equation}
%
\begin{equation}
 R_{\rm I}(r) = (2\kappa_{\rm I}r)^l e^{-\kappa_{\rm I}r} \cdot 
 [\tilde A_{\rm I} M(a_{\rm I}, 2l+2, 2\kappa_{\rm I}r) 
 + \tilde B_{\rm I} U(a_{\rm I}, 2l+2, 2\kappa_{\rm I}r)] \,\,.
\label{eq2.37p}
\end{equation}

We now turn to the continuity conditions which enable us to determine the 
amplitudes $A_{\rm I}$, $B_{\rm I}$ (or $\tilde A_{\rm I}$, $\tilde B_{\rm I}$),
$C_0$, and the energy eigenvalues $E$. Once these parameters will be obtained, 
we shall normalize the entire radial function to 1.

The radial wave function $R(r)$ and its 1$^{\rm st}$ derivatives must be 
continuous at the limits $r = R_2$ and $r = R_1$ of the regions II and I:
\begin{equation}
 R_{\rm II}(R_2) = R_{\rm I}(R_2) \,\,,
\label{eq2.38}
\end{equation}
%
\begin{equation}
 {dR_{\rm II}(R_2)\over dR_2} = {dR_{\rm I}(R_2)\over dR_2} \,\,,
\label{eq2.39}
\end{equation}
%
\begin{equation}
 R_{\rm I}(R_1) = R_0(R_1)  \,\,,
\label{eq2.40}
\end{equation}
%
\begin{equation} 
 {dR_{\rm I}(R_1)\over dR_1} = {dR_0(R_1)\over dR_1} \,\,.
\label{eq2.41}
\end{equation}
We proceed as follows:\\
Eqs. (\ref{eq2.38}) and (\ref{eq2.39}) are 2 linear inhomogeneous equations for 
the amplitudes $A_{\rm I}$, $B_{\rm I}$ (or $\tilde A_{\rm I}$, 
$\tilde B_{\rm I}$).
We solve them for these amplitudes and substitute the result into the equation
\begin{equation}
 {d\ln R_{\rm I}(R_1)\over dR_1} = {d\ln R_0(R_1)\over dR_1} \,\,,
\label{eq2.42}
\end{equation}
i.e. the ratio of the Eqs. (\ref{eq2.41}) and (\ref{eq2.40}).
Eq.~(\ref{eq2.42})  represents a transcendental equation for the energy. It is
solved numerically. The amplitude $C_0$ can subsequently be obtained from
(\ref{eq2.40}). The resulting wavefunction $R(r)$ is not yet normalized to 1.
This can be easily achieved by the replacement
\begin{equation} 
 R(r) \rightarrow \tilde R(r): = {\cal N} \cdot R(r) \,\,,
\label{eq2.43}
\end{equation}
with
\begin{equation}
 {\cal N} = \left\{\int\limits^\infty_0 dr r^2 R^2(r)\right\}^{-{1\over 2}}\,.
\label{eq2.44}
\end{equation}

Let us present this procedure in more detail:\\
We introduce the abbreviations:
\begin{equation}
 f^{(1)}_l(\kappa_{\rm I}r) \equiv [f^{(1)}_l(x)]_{x=\kappa_{\rm I}r} = 
 \left[\sqrt{{\pi\over 2x}} I_{l+{1\over 2}}(x)\right]_{x=\kappa_{\rm I}r} \,\,,
\label{eq2.45.1}
\end{equation}
%
\begin{equation}
 f^{(2)}_l(\kappa_{\rm I}r) \equiv [f^{(2)}_l(x)]_{x=\kappa_{\rm I}r}: = 
 \left[\sqrt{{\pi\over 2x}} K_{l+{1\over 2}}(x)\right]_{x=\kappa_{\rm I}r} \,\,,
\label{eq2.45.2}
\end{equation}
%
\begin{equation} 
 \tilde f^{(1)}_l(2\kappa_{\rm I}r) 
 \equiv [\tilde f^{(1)}_l(y)]_{y=2\kappa_{\rm I}r}: = 
 \left[y^l e^{-{y\over 2}} M(a_{\rm I}, 2l+2,y)\right]_{y=2\kappa_{\rm I}r}\,\,,
\label{eq2.46.1}
\end{equation}
%
\begin{equation}
 \tilde f^{(2)}_l(2\kappa_{\rm I}r) 
 \equiv [\tilde f^{(2)}_l(y)]_{y=2\kappa_{\rm I}r}: = 
 \left[y^l e^{-{y\over 2}} U(a_{\rm I}, 2l+2,y)\right]_{y=2\kappa_{\rm I}r}\,\,.
\label{eq2.46.2}
\end{equation}
Furthermore, we use the prime ' to denote the derivative with respect to the
variable given as argument:
$$
 f_l^{(i)}\,'(\kappa_{\rm I}R_2): = 
   \left[{df^{(i)}_l(x)\over dx}\right]_{x=\kappa_{\rm I}R_2} \,\,,
$$
$$
 \tilde f_l^{(i)}\,'(2\kappa_{\rm I}R_2): = 
   \left[{d\tilde f^{(i)}_l(y)\over dy}\right]_{y=2\kappa_{\rm I}R_2} \,\,,
$$
$$
 j_l'(\kappa_{\rm II}R_2):=\left[{dj_l(x)\over dx}\right]_{x=\kappa_{\rm II}
  R_2} \,\,,
$$
a.s.o.\\
From the continuity equations (\ref{eq2.38}), (\ref{eq2.39}) we find the 
amplitudes $A_{\rm I}$, $B_{\rm I}$ and $\tilde A_{\rm I}$, $\tilde B_{\rm I}$ 
to be given by very similar relations
\begin{equation}
 A_{\rm I} = {j_l(\kappa_{\rm II}R_2) f^{(2)}_l\,'(\kappa_{\rm I}R_2) - 
  {\kappa_{\rm II}\over \kappa_{\rm I}} f^{(2)}_l(\kappa_{\rm I}R_2) 
  j'_l(\kappa_{\rm II}
  R_2) \over [W(f^{(1)}_l, f^{(2)}_l)]_{\kappa_{\rm I}R_2}} \,\,,
\label{eq2.47.1}
\end{equation}
%
\begin{equation}
  B_{\rm I} = -{j_l(\kappa_{\rm II}R_2) f^{(1)}_l\,'(\kappa_{\rm I}R_2) - 
  {\kappa_{\rm II}\over \kappa_{\rm I}} f^{(1)}_l(\kappa_{\rm I}R_2) 
  j'_l(\kappa_{\rm II}
  R_2) \over [W(f^{(1)}_l(x), f^{(2)}_l(x))]_{x=\kappa_{\rm I}R_2}} \,\,.
\label{eq2.47.2}
\end{equation}
The Wronskian in the denominators is given by (see Ref.~\cite{Abram}, 
Eq.~(10.2.8))
\begin{equation}
 W(f_l^{(1)}(x), f^{(2)}_l(x)): = f^{(1)}_l(x) f^{(2)}_l\,'(x) - f^{(2)}_l(x)
 f^{(1)}_l\,'(x) = -{\pi\over 2x^2} \,\,,
\label{eq2.47.3}
\end{equation}
%
\begin{equation}
 \tilde A_{\rm I} = {j_l(\kappa_{\rm II}R_2) \tilde f^{(2)}_l\,'
 (2\kappa_{\rm I}R_2) -
 {\kappa_{\rm II}\over 2\kappa_{\rm I}} \tilde f^{(2)}_l(2\kappa_{\rm I}R_2)
 j'_l(\kappa_{\rm II}R_2)\over [W(\tilde f^{(1)}_l(y), 
 \tilde f^{(2)}_l(y))]_{y=2\kappa_{\rm I}R_2}}  \,\,,
\label{eq2.48.1}
\end{equation}
%
\begin{equation}
 \tilde B_{\rm I} = -{j_l(\kappa_{\rm II}R_2) \tilde f^{(1)}_l\,'
 (2\kappa_{\rm I}R_2) -
 {\kappa_{\rm II}\over 2\kappa_{\rm I}} \tilde f^{(1)}_l(2\kappa_{\rm I}R_2)
 j'_l(\kappa_{\rm II}R_2)\over [W(\tilde f^{(1)}_l(y), 
 \tilde f^{(2)}_l(y))]_{y=2\kappa_{\rm I}R_2}} \,\,.
\label{eq2.48.2}
\end{equation}
The Wronskian in the denominators of (\ref{eq2.48.1}) and (\ref{eq2.48.2}) 
can be related to the Wronskian $W(M(a_{\rm I},2l+2,y), U(a_{\rm I}, 2l+2,y))$:
\begin{equation}
\begin{array}{l}
  W(\tilde f^{(1)}_l(y), \tilde f^{(2)}_l(y)): = \tilde f^{(1)}_l(y) 
 {\tilde f^{(2)}_l\,'(y)} - \tilde f^{(2)}_l(y)  {\tilde f^{(1)}_l\,'(y)} 
 \\
 W(\tilde f^{(1)}_l(y), \tilde f^{(2)}_l(y) = y^{2l} e^{-y} \cdot 
 W(M(a_{\rm I},2l+2,y), U(a_{\rm I},2l+2,y))
\end{array}
\label{2.48.3}
\end{equation}
Using the differential Eq.~(\ref{eq2.34}), one finds for $W(M,U)$ 
(see Ref.~\cite{Abram}, Eq.~(13.1.22))
\begin{equation}
 W(M(a_{\rm I},2l+2,y), U(a_{\rm I}, 2l+2,y)) = -{\Gamma(2l+2) \cdot e^y \over 
 \Gamma(a_{\rm I}) \cdot y^{2l+2}}
\label{eq2.48.4} 
\end{equation}
and thus for $W(\tilde f^{(1)}_l(y), \tilde f^{(2)}_l(y))$:
\begin{equation}
 W(\tilde f^{(1)}_l(y), \tilde f^{(2)}_l(y)) = -{\Gamma(2l + 2)\over 
 \Gamma(a_{\rm I}) \cdot y^2}\,.
\label{eq2.48.5}
\end{equation}
Substituting (\ref{eq2.47.3}) into (\ref{eq2.47.1}) and (\ref{eq2.47.2}) and, 
analogously, substituting (\ref{eq2.48.5}) into (\ref{eq2.48.1}) and
(\ref{eq2.48.2}) we obtain the following forms of the amplitudes in region I:
\begin{equation}
 A_{\rm I} =  -{2\kappa^2_{\rm I}R^2_2\over \pi} \left[j_l(\kappa_{\rm II}R_2)
 f^{(2)}_l\,' (\kappa_{\rm I}R_2) - {\kappa_{\rm II}\over \kappa_{\rm I}} 
 f^{(2)}_l(\kappa_{\rm I}R_2) j'_l(\kappa_{\rm II}R_2)\right]  \,\,,
\label{eq2.49.1}
\end{equation}
%
\begin{equation}
 B_{\rm I} =  {2\kappa^2_{\rm I}R^2_2\over \pi} \left[j_l 
 (\kappa_{\rm II}R_2)
 f^{(1)}_l\,' (\kappa_{\rm I}R_2) - {\kappa_{\rm II}\over \kappa_{\rm I}} 
 f^{(1)}_l(\kappa_{\rm I}R_2) j'_l(\kappa_{\rm II}R_2)\right] \,\,,
\label{eq2.49.2}
\end{equation}
%
\begin{equation}
 \tilde A_{\rm I} = -{4\kappa^2_{\rm I} R^2_2\Gamma(a_{\rm I})\over 
 \Gamma(2l+2)} \left[j_l(\kappa_{\rm II}R_2) \tilde f^{(2))}_l\,'(2\kappa_{\rm I}
 R_2) - {\kappa_{\rm II}\over 2\kappa_{\rm I}} \tilde f^{(2)}_l(2\kappa_{\rm I}R_2) 
 j_l'(\kappa_{\rm II}R_2)\right] \,\,,
\label{eq2.50.1}
\end{equation}
%
\begin{equation}
 \tilde B_{\rm I} = {4\kappa^2_{\rm I} R^2_2\Gamma(a_{\rm I})\over 
 \Gamma(2l+2)} \left[j_l(\kappa_{\rm II}R_2) \tilde f^{(1))}_l\,'(2\kappa_{\rm I}
 R_2) - {\kappa_{\rm II}\over 2\kappa_{\rm I}} \tilde f^{(1)}_l(2\kappa_{\rm I}R_2) 
 j_l'(\kappa_{\rm II}R_2)\right] \,\,.
\label{eq2.50.2}
\end{equation}
With (\ref{eq2.49.1}), (\ref{eq2.49.2}) or (\ref{eq2.50.1}), (\ref{eq2.50.2}) 
the amplitudes of the 2 linearly independent solutions in region I are 
represented as a function of the energy $E$. The eigenvalues can now be 
obtained from the ratio of the two continuity conditions (\ref{eq2.41}) and 
(\ref{eq2.40}), i.e. from the equation
\begin{equation}
 \left({d\ln R_{\rm I}(r)\over dr}\right)_{r=R_1} = \left({d\ln R_0(r) 
 \over dr}\right)_{r=R_1} \,.
\label{eq2.51}
\end{equation}
This is a transcendental equation for the eigenenergies. It has infinitely many
solutions close to the energy $E = 0$ due to the ''infinite range'' of the 
Coulomb potential. In spite of the complicated nature of the wavefunctions it 
is not difficult to find the numerical solutions of (\ref{eq2.51}). More 
explicitly, the logarithmic derivatives in (\ref{eq2.51}) look as follows:
\begin{equation}
 \left[{d\ln R_0(r)\over dr}\right]_{r=R_1} = \left[{{d\over dr}((2\lambda r)^l
 U(a_0, 2l+2, 2\lambda r) e^{-\lambda r}) \over (2\lambda r)^l U(a_0, 2l+2,
 2\lambda r) e^{-\lambda r}}\right]_{r=R_1}
\label{eq2.52}
\end{equation}
%
\begin{equation}
 \left[{d\ln R_{\rm I}(r)\over dr}\right]_{r=R_1} = \left[{{d\over dr}
 (A_{\rm I}f^{(1)}_l(r) + B_{\rm I} f^{(2))}_l(r)) \over A_{\rm I} f^{(1)}_l(r)
 + B_{\rm I} f^{(2)}_l(r)}\right]_{r=R_1}
\label{eq2.53.1}
\end{equation}
%
\begin{equation}
 \left[{d\ln R_{\rm I}(r)\over dr}\right]_{r=R_1} = \left[{{d\over dr}
 (\tilde A_{\rm I}\tilde f^{(1)}_l(r) + \tilde B_{\rm I}\tilde f^{(2))}_l(r)) 
 \over \tilde A_{\rm I} \tilde f^{(1)}_l(r)
 + \tilde B_{\rm I} \tilde f^{(2)}_l(r)}\right]_{r=R_1}
\label{eq2.53.2}
\end{equation}
Eq. (\ref{eq2.53.2}) and (\ref{eq2.53.1}) deal with the case of the exact 
potential and of the approximate constant potential in region I. In calculating
the derivative  $f^{(i)}_l\,'(r)$ or $\tilde f^{(i)}_l\,'(r)$ one can use
properties of the  different functions involved.

Having determined the eigenvalues from Eq.~(\ref{eq2.51}) one obtains the 
amplitude $C_0$ from Eq.~(\ref{eq2.40}) and thereafter the correct 
normalization of the entire function must be achieved by numerical calculation
of the normalization constant (\ref{eq2.44}).


\section{Polarization effects in a macroscopic description}
\label{sec:3}

In Sec.~2, only the polarization of the insulator which is
produced by a constant positive surface charge density
$Ze_0/(4\pi R^2_2)$ on the surface $S_2$ of the metal was
taken into account.  However, the charge distribution
$\rho_e(\vec r)$ of the externally localized electrons produces
not only additional polarization effects in the insulating
surface layer but also modifies the distribution of the positive
surplus charge on the surface $S_2$ of the metal.  In this
section, we present a simple macroscopic description of the
polarization effects which is based on Maxwell's equations in
the presence of homogeneous dielectric media.

Outside of the metallic core, i.e. in the region I ($R_2 < r <
R_1$) of the insulator, and in the vacuum 0 ($r > R_1$)
surrounding the insulator the ''electric field'' $\vec E$ and
the ''displacement'' $\vec D$ are related by:
\begin{eqnarray}
 &\vec D(\vec r) = \varepsilon_{\rm I} \vec E(\vec r) = (1 + 4\pi \chi_{\rm I})
 \vec E(\vec r) & {\rm for}~R_2 < r < R_1~~~{\rm region~ I} \,\,,
\label{eq72}\\
 &\vec D(\vec r) = \vec E(\vec r) & {\rm for}~r>R_1~~~~{\rm region~ 0}~~~~,
\label{eq73}
\end{eqnarray}
where $\varepsilon_{\rm I}$ and $\chi_{\rm I}$ are the
''dielectric constant'' and the ''electric susceptibility'' of
the insulator, respectively. The fields satisfy Maxwell's
equations
\begin{equation}
 div~\vec D = 4\pi\rho_e(\vec r) \,\,,
\label{eq74}
\end{equation}
\begin{equation}
rot~\vec E = 0  \,\,.
\label{eq75}
\end{equation}
For the sake of simplicity, we assume that the distribution
$\rho_e(\vec r)$ of the externally located electrons and, as a
consequence, also the positive surface charge distribution
$\rho_{S_2}(\vec r)$ on $S_2$, are \underline{azimuthally symmetric}, i.e.
that they depend only on the polar angle $\vartheta$ and the
radial coordinate $r$. They can thus be written in the form
\begin{eqnarray}
&& \rho_e(\vec r) = \rho_e(r,\vartheta) = \sum^\infty_{l=0} \rho_l(r)\, Y_{l0}
   (\vartheta)\,\,, 
\label{eq76}\\
&& \rho_{S_2}(\vec r) = \rho_{S_2} (R_2,\vartheta) = \sum^\infty_{l=0}
 \rho^{S_2}_l \, Y_{l0}(\vartheta) \,\,.
\label{eq77}
\end{eqnarray}
Integrating the surface charge density over the surface $S_2$ must 
yield the total positive surface charge $Ze_0$. This implies the 
condition
\begin{equation}
 Ze_0 = \int d^3r \, \delta(r - R_2)\, \rho_{S_2}(R_2,\vartheta)
\label{eq78}
\end{equation}
or, with (\ref{eq77})
\begin{equation}
 \rho^{S_2}_{l=0} = {Ze_0\over \sqrt{4\pi}R^2_2}\,.
\label{eq79}
\end{equation}
The components $\rho_l(r)$ of the electronic charge distribution 
are related  to the total electronic charge distribution by
\begin{equation}
 \rho_l(r) = 2\pi \int\limits^\pi_0 d\vartheta\sin\vartheta\,\rho_e(r,\vartheta)
 Y_{l0}(\vartheta)\,.
\label{eq80}
\end{equation}
The Maxwell equations (\ref{eq74}) and (\ref{eq75}) are equivalent
to the Poisson equation
\begin{equation}
 \Delta\Phi_{\rm I}(r,\vartheta) = -{4\pi\over\varepsilon_{\rm I}}
 \rho_e(r,\vartheta)
\label{eq81}
\end{equation}
in region I and to
\begin{equation}
 \Delta\Phi_0(r,\vartheta) = -4\pi\rho_e(r,\vartheta)
\label{eq82}
\end{equation}
in region 0. Here, the potential
\begin{eqnarray}
 \Phi(r,\vartheta) = \Phi_{\rm I}(r,\vartheta) &~~ {\rm for}~R_2 < r < R_1 \,\,,
\label{eq83}\\
 \Phi(r,\vartheta) =  \Phi_0(r,\vartheta) &~~ {\rm for}~r > R_1 
\label{eq84}
\end{eqnarray}
is related to the electric field $\vec E$ by
\begin{equation}
 \vec E = -\vec\nabla\Phi\,.
\label{eq85}
\end{equation}
The following boundary conditions must be fulfilled at the surfaces $S_2$
and $S_1$:
\begin{equation}
 -\varepsilon_{\rm I} {\partial\Phi_{\rm I}(R_2,\vartheta)\over\partial R_2}
 = 4\pi\, \rho_{S_2}(R_2,\vartheta)
\label{eq86}
\end{equation}
%
\begin{equation}
 {\partial\Phi_{\rm I}(R_2,\vartheta)\over \partial\vartheta} = 0
\label{eq87}
\end{equation}
%
\begin{equation}
 \varepsilon_{\rm I} {\partial\Phi_{\rm I}(R_1,\vartheta) \over \partial R_1}
 = {\partial\Phi_0(R_1,\vartheta)\over\partial R_1}
\label{eq88}
\end{equation}
%
\begin{equation}
 \Phi_{\rm I}(R_1,\vartheta) = \Phi_0(R_1,\vartheta)\,.
\label{eq89}
\end{equation}
Furthermore, the potential $\Phi_0(r,\vartheta)$ must satisfy the 
asymptotic condition
\begin{equation}
 \Phi_0(r,\vartheta) \rightarrow 0 \left({1\over r}\right)\,.
\label{eq90}
\end{equation}
The conditions (\ref{eq86}) and (\ref{eq88}) are obtained from
the Maxwell equation (\ref{eq74}), and the conditions (\ref{eq87})
and (\ref{eq89}) result from (\ref{eq75}) and the requirement that
the classical electric field should be everywhere finite.

The solutions $\Phi_{\rm I}$ and $\Phi_0$ of the Poisson equations
(\ref{eq81}) and (\ref{eq82}) in the regions I and 0, respectively,
can be written in the form:
\begin{eqnarray}
&& \Phi_{\rm I}(r,\vartheta) = {1\over\varepsilon_{\rm I}} \, 
 \Psi(r,\vartheta) + \varphi_{\rm I}(r,\vartheta) \,\,,
\label{eq91}\\
&& \Phi_0(r,\vartheta) = \Psi(r,\vartheta) + \varphi_0(r,\vartheta)\,,
\label{eq92}
\end{eqnarray}
where $\Psi(r,\vartheta)$ is the Coulomb potential produced by the
electronic charge distribution $\rho_e(\vec r',\vartheta')$
\begin{equation}
 \Psi(r,\vartheta): = \int d^3r' {\rho_e(r',\vartheta') \over 
 |\vec r - \vec r'|}\,.
\label{eq93}
\end{equation}
The charge distribution $\rho_e(\vec r)$ itself is to be calculated
from the wavefunctions of the bound and predominantly externally 
located (i.e. in region 0, $r > R_1$) electrons. The ''conduction
electrons'' which are located internally (i.e. in region II, $r < R_2$) 
do not contribute to $\rho_e(\vec r)$ because they are already taken
into account by the number $Z$ of positive surplus charges. 
The density $\rho_e(\vec r)$ is thus mainly located in region 0 and
it is very small in region I, i.e. inside of the insulating surface layer, 
because the eigenenergies of the electrons are far below the positive barrier
in region I.

The function $\varphi_{\rm I}(r,\vartheta)$ is the general azimuthally symmetric
solution of the Laplace equation in region I
\begin{equation}
 \varphi_{\rm I}(r,\vartheta) = \sum^\infty_{l=0} \left(C^{\rm I}_l r^l + 
 {D^{\rm I}_l \over r^{l+1}}\right) Y_{l0}(\vartheta)
\label{eq94}
\end{equation}
and $\varphi_0(r,\vartheta)$ the corresponding most general solution of
the Laplace equation in region 0 satisfying in addition the asymptotic
condition (\ref{eq90}):
\begin{equation}
 \varphi_0(r,\vartheta) = \sum^\infty_{l=0} {D^0_l\over r^{l+1}} Y_{l0}(\vartheta)\,.
\label{eq95}
\end{equation}
The Coulomb potential $\Psi(r,\vartheta)$ can be written as
\begin{equation}
  \Psi(r,\vartheta) = \sum^\infty_{l=0} {4\pi\over 2l+1} \left(a_l(r) r^l + 
 {b_l(r)\over r^{l+1}}\right) Y_{l0}(\vartheta)\,,
\label{eq96}
\end{equation}
where the functions $a_l(r)$ and $b_l(r)$ are related to the components
$\rho_l(r)$ of the electronic charge distribution by 
\begin{eqnarray}
&& a_l(r): = \int\limits^\infty_r dr' \rho_l(r')\, r'^{1-l} \,\,,
\label{eq97} \\
&& b_l(r): = \int\limits_0^r dr' \rho_l(r')\, r'^{l+2} \,\,.
\label{eq98}
\end{eqnarray}
Substituting the explicit forms (\ref{eq94}) to (\ref{eq96}) into the
boundary conditions (\ref{eq86}), (\ref{eq88}), and (\ref{eq89}),
we obtain the relations
\begin{eqnarray}
&& -\sum^\infty_{l=0} \left\{{4\pi\over (2l+1)\varepsilon_{\rm I}} \left(a_l
 (R_2) lR^{l-1}_2 - {b_l(R_2)(l+1)\over R_2^{l+2}}\right) + C^{\rm I}_l 
 l R^{l-1}_2 \right. \nonumber\\
&&~~~~~~~~~~~~~\left. - {D^{\rm I}_l \cdot (l+1)\over R^{l+2}} + {4\pi\over 
\varepsilon_{\rm I}} \rho^{S_2}_l \right\} Y_{l0}(\vartheta) = 0 \,\,, 
\label{eq99} \\
&& \varepsilon_{\rm I} \sum^\infty_{l=0} \left\{C^{\rm I}_l l R^{l-1}_1 - 
 {D^{\rm I}_l \cdot (l+1)\over R^{l+2}_1} + {D^0_l \cdot (l+1) \over 
 R^{l+2}_1\varepsilon_{\rm I}}\right\} Y_{l0}(\vartheta) = 0 \,\,,
\label{eq100} \\
&&  \sum^\infty_{l=0} \left\{C^{\rm I}_l  R^l_1 +
 {D^{\rm I}_l \over R^{l+1}_1} - {D^0_l \over 
 R^{l+1}_1}\right\} Y_{l0}(\vartheta) = 0 \,\,.
\label{eq101}
\end{eqnarray}
Furthermore, using the identity (see for instance Ref.~\cite{Abram}) for 
$l\geq 1$
\begin{equation}
 \sin\vartheta{dY_{l0}\over d\vartheta} = {l(l+1)\over \sqrt{2l+1}}
 \left[{Y_{l+1,0}\over \sqrt{2l+3}} - {Y_{l-1,0}\over \sqrt{2l-1}}\right]
\label{eq102} 
\end{equation}
together with (\ref{eq91}), (\ref{eq92}) and (\ref{eq94}), the condition
(\ref{eq87}) can be shown to be equivalent to the relation
\begin{equation}
 {4\pi\over (2l+1)\varepsilon_{\rm I}} \left(a_l(R_2)R^l_2 + {b_l(R_2)\over 
 R^{l+1}_2}\right) + C^{\rm I}_l R^l_2 + {D^{\rm I}_l \over R^{l+1}_2} = 0\,\,,
\label{eq103}
\end{equation}
where $l \geq 1$. Eq. (\ref{eq103}) is seen to imply that the potential $\Phi_{\rm I}
(r,\vartheta)$ is constant on the surface $S_2$. For $l = 0$, we obtain from 
(\ref{eq99})
\begin{equation}
 D^{\rm I}_0 = {4\pi\over\varepsilon_{\rm I}} [-b_0(R_2) + R^2_2\rho^{S_2}_0]
\label{eq104}
\end{equation}
and from (\ref{eq100}) and (\ref{eq101}) the relations
\begin{eqnarray}
&& \varepsilon_{\rm I} D^{\rm I}_0 = D^0_0 \,\,, 
\label{eq105}\\
&& C^{\rm I}_0 + {D^{\rm I}_0\over R_1} = {D^0_0\over R_1} \,\,.
\label{eq106}
\end{eqnarray}
Hence, with (\ref{eq104}), we find
\begin{equation}
 C^{\rm I}_0 = {(\varepsilon_{\rm I} - 1)4\pi\over \varepsilon_{\rm I}R_1}
 [-b_0(R_2) + R^2_2 \rho^{S_2}_0]
\label{eq107}
\end{equation}
and
\begin{equation}
 D^0_0 = 4\pi [-b_0(R_2) + R^2_2\rho^{S_2}_0] \,\,.
\label{eq108}
\end{equation}
For $l \geq 1$, we obtain from (\ref{eq99}) and (\ref{eq103}):
\begin{eqnarray}
&& C^{\rm I}_l =  -{4\pi\over (2l+1)\varepsilon_{\rm I}} \left[a_l(R_2) + 
 {\rho^{S_2}_l\over R^{l-1}_2}\right] \,\,,
\label{eq109}\\
&&  D^{\rm I}_l =  {4\pi\over (2l+1)\varepsilon_{\rm I}} \left[-b_l(R_2) + 
 \rho^{S_2}_l R^{l+1}_2\right] \,\,.
\label{eq110}
\end{eqnarray}
For the coefficients $D^0_l$, we find the relations
\begin{equation}
 D^0_l = C^{\rm I}_l R^{2l+1}_1 + D^{\rm I}_l 
\label{eq111}
\end{equation}
and
\begin{equation}
 D^0_l = \varepsilon_I \left[-C^{\rm I}_l \cdot {l\over (l+1)} R^{2l+1}_1
 + D^{\rm I}_l\right]
\label{eq112}
\end{equation}
from the Eqs. (\ref{eq101}) and (\ref{eq100}), which hold for $l \geq 0$.

The equality of the righthand sides of (\ref{eq111}) and (\ref{eq112}) 
together with (\ref{eq109}) and (\ref{eq110}) yields a condition for the
surface charge components $\rho^{S_2}_l$ for $l \geq 1$. From this equation
one finds the explicit form
\begin{equation}
 \rho^{S_2}_l = R^{l-1}_2 { \left\{-a_l(R_2)\left({l\over l+1}\varepsilon_{\rm I}
 + 1\right) + b_l(R_2) R^{-(2l+1)}_1 \cdot (\varepsilon_{\rm I} - 1)\right\} \over
 \left\{\left({l\over l+1} \varepsilon_{\rm I} + 1\right) + \left({R_2\over 
 R_1}\right)^{2l+1} (\varepsilon_{\rm I} - 1)\right\}} 
\label{eq113} 
\end{equation}
for $l \geq 1$. The component $\rho^{S_2}_0$ is determined by the conservation
of the total positive charge on $S_2$ and is given by Eq. (\ref{eq79}).

For $l \geq 1$, one obtains from (\ref{eq109}), (\ref{eq110}) and 
(\ref{eq112})
\begin{equation}
 D^0_l = {4\pi\over (2l+1)} \left\{{l\over l+1} R^{2l+1}_1 a_l(R_2) - b_l(R_2) + 
 \rho^{S_2}_l \cdot \left({l\over l+1} {R^{2l+1}_1\over R^{l-1}_2} + 
 R^{l+2}_2\right)\right\} 
\label{eq114}
\end{equation}
As we have mentioned already, the charge distribution $\rho_e(r,\vartheta)$
of the externally located electrons is vanishingly small for $r < R_1$.
Consequently, we have 
\begin{equation}
 b_l(r) \approx 0~~~~~~{\rm for}~~~~~~r < R_1
\label{eq115}
\end{equation}
and 
\begin{equation}
 a_l(r) \approx a_l(R_1)~~~~{\rm for}~~~~r < R_1 \,\,.
\label{eq116}
\end{equation}

The potentials $\Phi_0(\vec r)$, $\Phi_{\rm I}(\vec r)$, 
$\Phi_{\rm II} = \Phi_{\rm I}(R_2)$, and the charge distribution $\rho_{S_2}
(R_2,\vartheta)$ on the surface $S_2$ are thus fully determined for a given
azimuthally symmetric distribution $\rho_e(r,\vartheta)$ of the externally
located electrons.

The potential $\tilde V(\vec r)$ acting on an electron in our ''mesoscopic 
atom'' is related to the potential functions as follows:
\begin{eqnarray}
 &\tilde V(r,\vartheta) = 
 -e_0\Phi_{\rm II} - V^0_{\rm II} = -e_0\Phi_{\rm I}(R_2) - V^0_{\rm II} 
 & ~~{\rm for}~~r < R_2 
\label{eq117} \,\,,\\
 &\tilde V(r,\vartheta) = 
 -e_0\Phi_{\rm I} (r,\vartheta) + B_0 &~~{\rm for}~~R_2 < r < R_1
\label{eq118} \,\,,\\
 &\tilde V(r,\vartheta) = 
 -e_0\Phi_0(r,\vartheta) & ~~{\rm for}~~r > R_1~(\rm region~ 0) \,\,.
\label{eq119}
\end{eqnarray}
The potential $(-V^0_{\rm II})$ represents the interaction of a conduction
electron with the average ionic background charge in the metal and has to be
added to the constant potential $-e_0\Phi_{\rm II}$ produced by the positive
surplus charge on $S_2$. The potential barrier $B_0$ to be added in the
insulating surface layer is related to the average energy gap between the
electronic bands. Since it is of quantum mechanical origin, it is not contained
in the classically determined term $-e_0\Phi_{\rm I} (r,\vartheta)$.

The Schr\"odinger equation with the potential $\tilde V(r,\vartheta)$ can no 
longer be solved analytically. One could diagonalize the Hamiltonian
\begin{equation}
 \widehat H = -{\hbar^2\over 2M}\Delta + \tilde V(r,\vartheta)
\label{eq120}
\end{equation}
within the set of analytical eigenfunctions which we have determined in 
Sec.~2. This represents a selfconsistency problem, because the 
eigenfunctions of the Hamiltonian (\ref{eq120}) depend on the charge
distribution $\rho_e(\vec r)$ which is calculated from them. 
The solution thus requires an iteration process.

Let us consider two special cases for which the results of this section 
simplify considerably:
\begin{itemize}
\item[1.] If we neglect the polarization effects produced by the external 
electrons, that is if we put
\begin{equation}
 \rho_e(\vec r) = 0\,,
\label{eq121}
\end{equation}
the quantities $a_l(r)$, $b_l(r)$ vanish for $l \geq 0$, and, as a consequence,
so do the components $\rho^{S_2}_{l\geq 1}$. The positive surplus charge
density on $S_2$ is then spherically symmetric and given by
\begin{equation}
 \rho_{S_2}(r,\vartheta) = \rho^{S_2}_0 Y_{00} = {Ze_0\over 4\pi R^2_2}\,.
\label{eq122}
\end{equation}
It is easily seen that, in this case, the potential $V(r,\vartheta)$ becomes
equal to the potential $V(r)$ we considered in Sec.~2.

\item[2.] If we assume that the charge density $\rho_e(\vec r)$ of the 
external electrons is spherically symmetric
\begin{equation}
\rho_e(\vec r) = \rho_e(r)\,,
\label{eq123}
\end{equation}
only the component $\rho_{l=0}(r)$ in (\ref{eq76}) is unequal 0
\begin{equation}
 \rho_l(r) = \delta_{l0} \cdot \sqrt{4\pi}\rho_e(r)
\label{eq124}
\end{equation}
and, consequently, the quantities $a_l(r)$, $b_l(r)$, and $\rho^{S_2}_l$ 
vanish for $l \geq 1$. In this case, the potentials $\Phi_{\rm II}$,
$\Phi_{\rm I}$ and $\Phi_0$ take the following form
\begin{eqnarray}
&& \Phi_{\rm II} = \Phi_{\rm I}(R_2)~~~~({\rm for}~~r < R_2) \,\,,
\label{eq125} \\
&& \Phi_{\rm I}(r) = \left(a_0(r) + {b_0(r)\over r}\right) \sqrt{4\pi}
 + {(\varepsilon_{\rm I} - 1)\over \varepsilon_{\rm I}} {Ze_0\over R_1} + 
 {Ze_0\over \varepsilon_{\rm I}r}~~~({\rm for}~~R_2 < r < R_1) \,\,,
\label{eq126} \\
&& \Phi_0(r) = \left(a_0(r) + {b_0(r)\over r}\right) \sqrt{4\pi}
 + {Ze_0\over r}~~~~({\rm for}~~r > R_1) \,\,.
\label{eq127}
\end{eqnarray}
\end{itemize}

The case of a spherically symmetric distribution of the
externally located electrons is realistic at finite temperature,
because magnetic substates of a given electronic level are
expected to be occupied with the same probability whenever the
mesoscopic atom is in contact with a thermal reservoir of finite
temperature.

The functions $a_0(r)$ and $b_0(r)$ depend themselves on the
electronic distribution.  For $r \gg R_1$, $a_0(r)$ decreases to
zero as one can see from (\ref{eq97}). The quantity
$\sqrt{4\pi}b_0(r)$ represents the total charge of the
externally located electrons inside the sphere of radius $r$
\begin{equation}
 Q_e(r): = \int\limits^r_0 dr'r'^2\int d\Omega' \rho_e(r',\vartheta') \,\,.
\label{eq128}
\end{equation}
Using (\ref{eq76}) in (\ref{eq128}) we find
\begin{equation}
 \sqrt{4\pi} b_0(r) = Q_e(r)\,.
\label{eq129}
\end{equation}
For $r >> R_1$, $\sqrt{4\pi} b_0(r)$ tends to the total charge 
$Q_e(\infty)$ of the externally located electrons.

If the mesoscopic atom is neutral, i.e. if there are as many
externally bound electrons as there are positive surplus charges
on the surface $S_2$ of the metallic core, the term
$b_0(r)\sqrt{4\pi}/r$ in Eq. (\ref{eq127}) cancels the
potential $Ze_0/r$ for $r \gg R_1$, as it should.

Let us study the asymptotic behaviour of the function $a_0(r)$
and $b_0(r)$ for $r \rightarrow \infty$: For simplicity, let us
assume that only the lowest external electron state is occupied
by two electrons, one with intrinsic spin up and the other one
with spin down. The external state of lowest energy belongs
necessarily to orbital angular momentum ($l = 0$) and to zero
number of radial modes.  We assume that the wavefunctions we
determined in Sec.~2 are realistic for $r
\gg R_1$. 
In region 0, the electronic density distribution is then given by
\begin{equation}
 \rho_e(r) = (-e_0) \cdot 2R^2_{l=0} (r;E) \, Y^2_{00} = {(-e_0)\over
 2\pi} \left({u_{l=0}(2\lambda r)\over 2\lambda r}\right)^2
\label{eq130}
\end{equation}
or, substituting the explicit wavefunction (\ref{eq2.19}), by
\begin{equation}
 \rho_e(r) = {(-e_0)\over 2\pi} C^2_0{\cal N}^2 
             U^2\left({2-d_0\over 2}, 2, 2\lambda r\right) e^{-2\lambda r} \,\,.
\label{eq131}
\end{equation}
Here, ${\cal N}$ is the overall normalization constant introduced by 
Eq.~(\ref{eq2.44}).

For $z_0 \gg 1$, the function $U(a_0,b_0,Z_0)$ can be replaced by (see
Ref.~\cite{Abram}, Eq.~(13.5.2))
\begin{equation}
 U(a_0,b_0,z_0)_{z_0\gg 1} \approx {1\over z^{a_0}_0} \left\{1 +
  \sum^\infty_{n=1}
 {(a_0)_n (1 + a_0 - b_0)_n\over n!} \left({1\over -z_0}\right)^n\right\}\,\,.
\label{eq132} 
\end{equation}
Retaining only the 1$^{\rm st}$ term in (\ref{eq132}), we obtain  for the
asymptotic density ($2\lambda r \ll 1$)
\begin{equation}
 \rho_e(r) \approx {(-e_0) (C_0{\cal N})^2\over 2\pi} (2\lambda r)^{d_0-2} 
 e^{-2\lambda r}
\label{eq133}
\end{equation}
and for the function $a_0(r)$
\begin{equation}
 a_0(r) \approx \sqrt{4\pi} \int\limits^\infty_r dr' r' \rho_e(r') 
        = {(-e_0)(C_0{\cal N})^2 \over \sqrt{\pi}(2\lambda)^2} 
          \int\limits^\infty_{2\lambda r} dz_0 (z_0)^{d_0-1} e^{-z_0} 
\label{eq134}
\end{equation}
The integral in (\ref{eq134}) is the ''incomplete $\Gamma$-function'' 
(see Ref.~\cite{Abram}, Eq.~(6.5.3), $Re(a) > 0$)
\begin{equation}
 \Gamma(a,x): = \int\limits^\infty_x dt t^{a-1} e^{-t}
\label{eq135}
\end{equation}
with the asymptotic expansion (Ref.~\cite{Abram}, Eq.~(6.5.32)
\begin{equation}
 \Gamma(a,x) \approx x^{a-1} e^{-x} \left\{1 + {a - 1\over x} + {(a - 1)(a-2)
\over x^2} + \ldots \right\}\,.
\label{eq136}
\end{equation}
From (\ref{eq134})--(\ref{eq136}), we obtain $a_0(r)$ in the form
\begin{equation}
 a_0(r) \approx {(-e_0)(C_0 {\cal N})^2 \over\sqrt{\pi}(2\lambda)^2} 
 (2\lambda r)^{d_0-1} e^{-2\lambda r}
\label{eq137}
\end{equation} 
for $2\lambda r\geq 1$.

Of course, one could raise the accuracy of the result (\ref{eq137}) by 
incorporating one or two more terms in (\ref{eq132}) and (\ref{eq136}).
The important point, however, is that $a_0(r)$ tends to zero exponentially
for large values of $r \gg R_1$. This justifies the assumption that the
electronic charge density $\rho_e(r)$ in the asymptotic regime 
$r \gg R_1$ is correctly described by the wavefunctions of Sec.~2,
which do not contain the electronic polarization effects.

The function $b_0(r)$ can be written in the form (see (\ref{eq98}))
\begin{equation}
 b_0(r) = \sqrt{4\pi} \int\limits^\infty_0 dr'r'^2 \rho_e(r') - \sqrt{4\pi}
 \int\limits^\infty_r dr' r'^2\rho_e(r') \,\,,
\label{eq138}
\end{equation}
with
\begin{equation}
 \int\limits^\infty_0 dr'r'^2\rho_0(r') = \sqrt{4\pi} \int\limits^\infty_0
 dr'r'^2\rho_e(r') = {Q_e(\infty)\over \sqrt{4\pi}} \,\,.
\label{eq139}
\end{equation}
The 2$^{\rm nd}$ term one the r.h.s. of (\ref{eq138}) is easily seen to have
the approximate form
\begin{equation}
 \int\limits^\infty_0 dr'r'^2\rho_0(r') \approx {(-e_0)(C_0{\cal N})^2\over 
\sqrt{\pi}(2\lambda)^3} (2\lambda r)^{d_0} e^{-2\lambda r}
\label{eq140}
\end{equation}
for $2\lambda r\geq 1$.

Thus the function $b_0(r)$ is asymptotically $(2\lambda r\geq 1)$ given by
\begin{equation}
 b_0(r) = {Q_{el}(\infty)\over \sqrt{4\pi}} + {e_0(C_0{\cal N})^2\over 
\sqrt{\pi}(2\lambda)^3} (2\lambda r)^{d_0} \cdot e^{-2\lambda r}\,,
\label{eq141}
\end{equation}
which means that it approaches exponentially the asymptotic value 
$Q_{e}(\infty)/\sqrt{4\pi}$.

In the general case of a $\vartheta$-dependent electronic charge density
$\rho_e(r,\vartheta)$, the quantities $a_{l\neq 0}(r)$ and $b_{l\neq 0}(r)$
are different from zero and their asymptotic behaviour can be investigated
in an analogous way. One finds that $a_{l>0}(r)$ tends exponentially to 
zero and $b_{l>0}(r)$ exponentially to constant values as $r$ goes to infinity.
Therefore, at large values of $r$, the dominant part of the potential 
$\tilde V(r,\vartheta)$ is the potential $V(r)$ given by Eq. (\ref{eq2.5.1}). 
Thereby, our assumption that the wavefunctions determined in Sec.~2 are 
essentially correct at large values of $\lambda r$ is seen to be justified.

For non-asymptotic values of $r$, the electronic charge density and the
resulting functions $a_0(r)$ and $b_0(r)$ must be obtained from the 
eigenfunctions of the Hamiltonian (\ref{eq119}) which contains the 
potential $\tilde V(r,\vartheta)$ defined in (\ref{eq117}) to (\ref{eq119}).
Let us formulate the special case of a spherically symmetric charge 
density and the resulting spherically symmetric potential $V(r)$ in somewhat
more detail:

The electronic charge density at a finite temperature $T$ is given by
\begin{equation}
 \rho_e(\vec r) = -e_0\sum^\infty_{n=1} \sum^\infty_{l=0} \left\{{2\over 
\exp\left({E_{nl} - \mu\over T}\right) + 1}\right\} R^2_l(r;E_{nl})
 \sum^{l}_{m=-l} |Y_{lm}(\vartheta,\varphi)|^2
\label{eq142}
\end{equation} 
or, using the relation (see for instance Ref.~\cite{Jacks}, Eq.~(3.69))
\begin{equation}
 \sum^{+l}_{m=-l}|Y_{lm}(\vartheta,\varphi)|^2 = {(2l+1)\over 4\pi} 
\label{eq143}
\end{equation}
by
\begin{equation}
 \rho_e(r) = -e_0 \sum^\infty_{n=1} \sum^\infty_{l=0} {(2l+1)
     \over 2\pi} \cdot {1\over \left\{\exp\left({E_{nl} - \mu\over T}\right) 
     + 1\right\}} \cdot R^2_l(r; E_{nl})  \,\,.
\label{eq144}
\end{equation}
Here, $E_{nl}$ is the $n$-th eigenvalue for given angular momentum $l$ and 
$R_l(r;E_{nl})$ is the corresponding eigenfunction of the radial Schr\"odinger 
equation. The parameter $\mu$ represents the chemical potential which 
corresponds to the number of externally located electrons.\\
The potential $\tilde V(r,\vartheta)$ has the form
\begin{eqnarray}
\tilde V(r) &=& -e_0\Phi_{\rm I}(R_2) - V^0_{\rm II} ~~~~~~~~~~~~~~
              ~~~~~~~~~~~~~~~~~~~~~~~~~~~~~~~~~~{\rm for}~~r_2 < R_2  \,\,,
\label{eq145}\\
\tilde V(r) &=& -e_0\Phi_{\rm I}(r) + B_0 \nonumber\\ 
            &=& B_0 - e_0 \left\{{4\pi\over\varepsilon_{\rm I}} 
            \left(a_0(r) + {b_0(r)\over r}\right) 
             + C^{\rm I}_0 + {D^{\rm I}_0\over r}\right\} Y_{00} 
                                           ~~~~~~~~~{\rm for}~~R_2<r<R_1 \,\,,
\label{eq146}\\
\tilde V(r) &=& -e_0\Phi_0(r) = -e_0 \left\{4\pi\left(a_0(r) 
             + {b_0(r)\over r}\right) + {D^0_0\over r}\right\} Y_{00} 
                                  ~~~~~~{\rm for}~r > R_1   \,\,,
\label{147}
\end{eqnarray}
where $D^{\rm I}_0$, $C^{\rm I}_0$, and $D^0_0$ are given by the Eq. 
(\ref{eq104}), (\ref{eq107}), and (\ref{eq108}), resp. The functions
$a_0(r)$ and $b_0(r)$ are defined by (\ref{eq97}) and (\ref{eq98}),
where $\rho_0(r') = \sqrt{4\pi}\rho_e(r')$. The wavefunctions 
$R_l(r;E_{nl})$ are solutions of the radial Schr\"odinger equation with
$\tilde V(r)$ given by (\ref{eq145}) - (\ref{147})
\begin{equation}
  \left\{-{\hbar^2\over 2M} \left({d^2\over dr^2} + {2\over r} {d\over dr}
 - {l(l+1)\over r^2}\right) + \tilde V(r) - E_{nl} \right\}
 \tilde R_l(r;E_{nl})_l  \,\,.
\label{eq148}
\end{equation}
Since $\tilde V(r)$ depends itself on the functions $\tilde R_l(r;E_{nl})$, 
one has to solve (\ref{eq148}) by iteration. One way to find the solutions 
$\tilde R_l(r;E_{nl})$ of (\ref{eq148}) is to expand the solutions
$R_l(r;E_{nl})$ in terms of the radial eigenfunctions 
$R_l(r;\buildrel\circ\over E_0)$ defined
in Sec.~2, where the potential $V(r)$ is given by Eq.~(\ref{eq2.5}) and related
to $\tilde V(r)$ by 
\begin{equation}
 V(r) = \lim_{\rho_e(r)\rightarrow 0} \tilde V(r)
\label{eq149}
\end{equation}
%
\begin{equation}
 \tilde R_l(r;E_{nl}) = \sum_\nu c_\nu R_l(r;\buildrel\circ\over E_\nu)
\label{eq150}
\end{equation}
$\buildrel\circ\over E_\nu$ are the eigenvalues obtained in Sec.~2.  In
Eq.~(\ref{eq150}) we left away the basis functions $R_l(r, E)$  corresponding
to continuous eigenvalues $\buildrel\circ\over E$ of Eq.~(\ref{eq2.8}).
Substitution of (\ref{eq150})  into (\ref{eq148}) leads to a nonlinear system
of equations for the expansion coefficients $c_\nu$ which must be solved by
iteration.


\section{Results}
\label{sec:4}

We present results on the spectrum of eigenenergies, the
eigenfunctions, and on the electromagnetic transition rate
between externally and internally localized states for a range
of values of the parameters $R_1$, $R_2$, and $\varepsilon_I$. So
far, we have not taken into account the effects presented in
Sec.~3, i.e. the polarization of the insulating dielectric
material by externally localized electrons. The additional
potential produced by this polarization is expected to modify
the wave functions in the vicinity of the surface $S_1$, perhaps
even rather substantially. At large distances (say a couple
of {\AA}ngstr\"om outside of the surface $S_1$),
the additional polarization potential tends rapidly
to zero. Consequently, the wavefunctions presented in this
article are expected to be trustworthy at distances of a couple
of {\AA}ngstr\"om from the surface $S_1$.

\subsection{Eigenfunctions and eigenenergies}

We studied only the case that 
the Coulomb potential $-Ze_0^2/(r\varepsilon_{\rm I})$ within
the insulator ($R_2 \leq r \leq R_1$) is small compared to the
barrier height $B$ which is related to the gap parameter 
$B_0$ by Eq. (\ref{eq2.4}). This means that we neglected the Coulomb
potential in region I.

\begin{table}[h]
\caption[TT]{The energies $Ze^2_0/(R_2\varepsilon_{\rm I})$,
$Ze^2_0(\varepsilon_{\rm I} - 1)/(R_1\varepsilon_{\rm I})$, and the
barrier height $B$ as a function of the radii $R_1$ and $R_2$ and of the
dielectric constant $\varepsilon_{\rm I}$. Energies are given in meV and
radii in \AA. The parameter $B_0$ is chosen equal to 2500 meV or 4000 meV.
\label{tab1}}
\begin{center}\begin{small}
\begin{tabular}{|l|c|c|c|c|c|c|c|c|}
\hline
$R_2$~(\AA) & 20 & 20 & 20 & 10 & 10 & 10 & 3.5 & 3.5 \\
\hline
$R_1$~(\AA) & 28 & 28 & 28 & 18 & 18 & 18 & 10 & 10 \\
\hline
$\varepsilon_{\rm I}$ & 115 & 28 & 10 & 115 & 28 & 10 & 115 & 10 \\
 & (Ti0$_2$) & (Hg0) & (Al$_2$0$_3$) & (Ti0$_2$) & (Hg0) & (Al$_2$0$_3$) &
 (Ti0$_2$) & (Al$_2$0$_3$) \\
\hline
$Z$ & 1 & 1 & 1 & 1 & 1 & 1 & 1 & 1  \\
\hline
${Ze^2_0\over R_2\varepsilon_{\rm I}}$~(meV) 
    & 6.26 & 25.7 & 72 & 12.5 & 51.4 & 144 & 35.8 & 411 \\
\hline
${Ze^2_0(\varepsilon_{\rm I} - 1)\over R_1\varepsilon_{\rm I}}$~(meV) 
    & 509.8 & 495.9 & 462.9 & 793 & 771 & 720 & 1427 & 1296 \\
\hline
$B$~(meV) for $B_0 = 4000$ 
    & 3490 & 3504 & 3537 & 3207 & 3229 & 3280 & 2573 & 2704 \\
\hline
$B$~(meV) for $B_0 = 2500$ 
    & 1990 & 2004 & 2037 & 1707 & 1729 & 1780 & 1073 & 1204 \\
\hline
\end{tabular}
\end{small}\end{center}
\end{table}

In Table~1 we present the energies $Ze_0^2/(R_2\varepsilon_{\rm I})$,
$Ze_0^2(\varepsilon_{\rm I} - 1)/(R_1\varepsilon_{\rm I})$, and
the barrier height $B$ for typical values of the parameters $R_{1,2}$
and $\varepsilon_{\rm I}$. It is seen that neglecting the screened Coulomb
potential $Ze_0^2/(R_1\varepsilon_{\rm I})$ in region I as compared to $B$
is sometimes a very good approximation, but, not in general, especially
not for $Z > 1$ and for radii $R_{1,2} \leq 10$~\AA.

The eigenvalues were determined by a numerical solution of the eigenvalue
equation (\ref{eq2.51}) using the form of $d\ln R_{\rm I}(R_1)/dR_1$ 
as given by the Eq. (\ref{eq2.53.1}), (\ref{eq2.49.1}), and (\ref{eq2.49.2}).
Up to the common normalization factor ${\cal N}$ (see (\ref{eq2.44})), the radial
eigenfunctions $R(r; l, E)$ are then obtained from the equations (\ref{eq2.10}),
(\ref{eq2.17}), and (\ref{eq2.19}) in the regions II, I, and 0, resp.

In order to simplify the comparison of results which correspond to 
different choices of the parameters $R_{1,2}$, $B$, $V_{II}$ and $Z$, it is
advantageous to use the energy unit $Ze^2_0/R_1$, i.e. the value 
of the Coulomb potential at the outer surface $S_1$. The resulting 
dimensionless quantities are denoted by
\begin{eqnarray}
&& \varepsilon: = {E\over Ze^2_0/R_1} \,\,,
\label{eq1} \\
&& b: = {B \over Ze^2_0/R_1} \,\,,
\label{eq2} \\
&& v_{\rm II}: = {V_{\rm II}\over Ze^2_0/R_1} \,\,,
\label{eq3} \\
&& \varepsilon_{\rm max}: = {(V_{\rm II} + Ze^2_0/R_2)\over Ze^2_0/R_1} =
 {E_{\rm max}\over Ze^2_0/R_1} = v_{\rm II} + {R_1\over R_2} \,\,.
\label{eq4}
\end{eqnarray}
Furthermore, it is convenient to introduce the typical
length parameters $R_0$ and $R_{00}$ (= 0,264589~\AA):
\begin{equation}
 R_0: = {\hbar^2\over 2M e^2_0Z} \equiv {R_{00}\over Z} \,\,.
\label{eq5}
\end{equation}
It what follows we give some technical details concerning
the determination of the eigenvalues and eigenenergies.

In region I, due to our approximation, the radial eigenfunction $R_{\rm I}$ 
is given by Eq. (\ref{eq2.17}), i.e. as a linear superposition of the 
modified spherical Bessel functions $f^{(1)}_l(x)$ and $f^{(2)}_l(x)$
defined in (\ref{eq2.45.1}) and (\ref{eq2.45.2}) with $x \equiv \kappa_{\rm I}r$. 
We calculate these functions by representing them in the form (see 
Ref.~\cite{Abram}, Chapt. 10)
\begin{eqnarray}
&& f^{(1)}_l(x) =  {1\over 2x} \left[R\left(l + {1\over 2},-x\right) e^x +
   (-1)^l\, R\left(l + {1\over 2},x\right) e^{-x} \right]  \,\,,
\label{eq6} \\
&& f^{(2)}(x) = {\pi\over 2x} R\left(l + {1\over 2},x\right) e^{-x} 
\label{eq7}
\end{eqnarray}
as a linear superposition of exponentially rising and decreasing terms.
Here, the function $R\left(l + {1\over 2},x\right)$ is defined 
(see Ref.~\cite{Abram}, Eq.~(10.2.11)) as
\begin{eqnarray}
&& R\left(l + {1\over 2},x\right):= \sum^l_{k=0} \left(l + {1\over 2},k\right)
 (2x)^{-k} \,\,,
\label{eq8} \\
&& \left(l + {1\over 2}, k\right): = {(l + k)!\over k!(l - k)!}\,\,.
\nonumber
\end{eqnarray}
It turned out to be convenient to rewrite $R_{\rm I}(r)$ in the form
\begin{eqnarray}
 R_{\rm I}(r) &=& A'_{\rm I} g^{(1)}_l(x) e^{x-\kappa_{\rm I}R_2} + {1\over\pi}
\left[B'_{\rm I} + (-1)^l A'_{\rm I} e^{-2\kappa_{\rm II}R_2}\right]\cdot  
\nonumber \\
&& \cdot\, g^{(2)}(x) e^{-(x - \kappa_{\rm I}R_2)} \,\,,
\label{eq10}
\end{eqnarray}
where the functions $g_l^{(1,2)}(x)$ are related to $f^{(1,2)}_l(x)$ by
\begin{eqnarray}
&& g^{(1)}_l(x): = f^{(1)}_l(x) e^{-x} \,\,,
\label{eq11} \\
&& g^{(2)}_l(x): = f^{(2)}_l(x) e^x    \,\,.
\label{eq12}
\end{eqnarray}
 The amplitudes $A'_{\rm I}$, $B'_{\rm I}$ are connected with the amplitudes
$A_{\rm I}$ and $B_{\rm I}$ of Eq. (\ref{eq2.17}) by
\begin{eqnarray}
&& A'_{\rm I} = A_{\rm I} e^{\kappa_{\rm I}R_2}  \,\,,
\label{eq13} \\
&& B'_{\rm I} = [(-1)^{l+1} A_{\rm I} + \pi B_{\rm I}] e^{-\kappa_{\rm I}R_2}\,.
\label{eq14}
\end{eqnarray}
The advantage of using (\ref{eq10}) instead of (\ref{eq2.17}) is that the 
functions $g^{1,2)}_l(y)$ can be numerically determined more comfortably
than the functions $f^{(1,2)}_l(y)$ and that they have a simpler asymptotic
behaviour
\begin{eqnarray}
&& g^{(1)}_l(x) \rightarrow {1\over 2x} + O\left({1\over x^2}\right) \,\,,
\label{eq15} \\
&& g^{(2)}_l(x) \rightarrow {\pi\over 2x} + O\left({1\over x^2}\right) \,\,.
\label{eq16}
\end{eqnarray}
From the continuity conditions (\ref{eq2.38}), (\ref{eq2.39}) one obtains
$A'_{\rm I}$ and $B'_{\rm I}$ in the form
\begin{eqnarray}
A'_{\rm I} &=& -{2\over\pi} (\kappa_{\rm I}R_2)^2\left[dg^{(2)}_l(\kappa_{\rm I}
R_2) \cdot j_l(\kappa_{\rm II}R_2) - {\kappa_{\rm II}\over\kappa_{\rm I}}
 g^{(2)}_l (\kappa_{\rm I}R_2) j'_l (\kappa_{\rm II} R_2)\right]  \,\,,
\label{eq17} \\
 B'_{\rm I} & = & (-1)^{l+1} A'_{\rm I} e^{-2\kappa_{\rm I}R_2} + 
 2(\kappa_{\rm I}R_2)^2 \left[dg^{(1)}(\kappa_{\rm I}R_2) j_l(\kappa_{\rm II}R_2)
\right.\nonumber \\ 
 && \left. - {\kappa_{\rm II} \over \kappa_{\rm I}} g^{(1)}_l (\kappa_{\rm I} R_2) 
 j'_l(\kappa_{\rm II}R_2)\right] \,\,,
\label{eq18}
\end{eqnarray} 
where the function $dg^{(1,2)}(x)$ are defined by
\begin{eqnarray}
&& dg^{(1)}(x) = e^{-x} {f^{(1)}_l}'(x) \,\,,
\label{eq19} \\
&& dg^{(2)}(x) = e^{+x} {f^{(2)}_l}'(x) \,\,.
\label{eq20}
\end{eqnarray}
Then, the continuity condition (\ref{eq2.51}) assumes the explicit form
\begin{eqnarray}
F(\varepsilon): &=& U\left(l+1 - {d_0\over 2},2l+2, 2\lambda R_1\right)
 [l R_{\rm I} + \kappa_{\rm I}R_1\tilde R_{\rm I}]  \nonumber\\
&+&  2\lambda R_1 \cdot (l + 1 - {d_0\over 2})
 U\left(l + 2 - {d_0\over 2}, 2l+3, 2\lambda R_1\right) 
 \cdot R_{\rm I} \label{eq21} \\
&+& (\lambda R_1 - l) U\left(l + 1 - {d_0\over 2}, 2l + 2, 2\lambda R_1\right)
 \cdot R_{\rm I} = 0
\nonumber
\end{eqnarray}
with
\begin{equation}
 R_{\rm I}: = R_{\rm I}(R_1) \,\,,
\label{eq22}
\end{equation}
%
\begin{eqnarray}
\tilde R_{\rm I}: &=&  A'_{\rm I}g^{(1)}_{l+1} (\kappa_{\rm I}R_1) 
 e^{\kappa_{\rm I}(R_1-R_2)} - {1\over \pi} B'_{\rm I} g^{(2)}_{l+1}
 (\kappa_{\rm I}R_1) e^{-\kappa_{\rm I}(R_1 - R_2)}\nonumber\\
 && - {(-1)^{l+1}\over\pi} A'_{\rm I} g^{(2)}_{l+1} (\kappa_{\rm I}R_1) 
 e^{-\kappa_{\rm I}(R_1-R_2)} e^{-2\kappa_{\rm I}R_2}  \,\,.
\label{eq23}
\end{eqnarray}
In terms of the dimensionless parameters (\ref{eq1}) to (\ref{eq4}) 
the arguments which appear in the transcendental equation (\ref{eq21}) 
take the form
\begin{eqnarray}
&& \kappa_{\rm I}R_1 = \sqrt{{R_1\over R_0}} \sqrt{b + |\varepsilon|} \,\,,
\label{eq24} \\
&& \kappa_{\rm II}R_2 = \sqrt{{R_2\over R_0}} \sqrt{{R_2\over R_1}}
  \sqrt{b + |\varepsilon|} = \sqrt{{R_2\over R_0}} \sqrt{{R_2\over R_1}}
   \sqrt{\varepsilon_{\rm max} - |\varepsilon|} \,\,,
\label{eq25} \\
&& \lambda R_1 = \sqrt{{R_1\over R_0}} \sqrt{\varepsilon} \,\,,
\label{eq26} \\
&& d_0 = \sqrt{{R_1\over R_0}} {1\over\sqrt{\varepsilon}} \,\,.
\label{eq27}
\end{eqnarray}
We determined the zeros of the function $F(\varepsilon)$ by calculating
numerically $F(\varepsilon)$ in small steps of $\varepsilon$ between limits
$\varepsilon_{\rm min}$ and $\varepsilon_{\rm max}$. Whenever $F(\varepsilon)$
changed sign from a given step to the next one, we refined the mesh by usually
a factor of 100 so as to determine the zero with the desired accuracy. We note 
that, for each given value of the angular momentum $l$, there is an infinity 
of eigenvalues as $\varepsilon$ approaches $\varepsilon = 0$ from below.
This is so as a result of the infinite range of the Coulomb potential in 
region 0. In the domain of small $|\varepsilon|$, i.e. of very weakly bound
electrons, and also for large well depth $v_{\rm II}$, the function 
$F(\varepsilon)$ oscillates between large positive and negative values.
Nevertheless, for a sufficiently narrow system of mesh points, the function
$F(\varepsilon)$ is of reasonable size for the meshpoint closest to a zero.
Therefore, we encountered no problems of numerical accuracy in determining
the eigenvalues.

Once an eigenvalue is determined, the amplitudes $A'_{\rm I}$, $B'_{\rm I}$ 
are obtained from (\ref{eq17}), (\ref{eq18}) then $A_{\rm I}$, $B_{\rm I}$
from (\ref{eq13}), (\ref{eq14}). The amplitude $C_0$ of the radial 
wavefunction $R_0(r)$ in region 0 (see Eq. (\ref{eq2.30}) is found from the
continuity condition (\ref{eq2.40}) and, finally, the overall normalisation
constant ${\cal N}$ from (\ref{eq2.44}). The integral in Eq. (\ref{eq2.44}) is
done numerically.

We note that the numerical determination of the eigenvalues and eigenfunctions
turned out to be extremely rapid. Thus it is conceivable that one uses these
eigenfunctions (in principle completed by the scattering states) as basis
functions for diagonalizing the additional polarization potentials studied in
Sec.~3.

We note in passing that $A'_{\rm I}$ has in general a very small value which 
is difficult to obtain with sufficient accuracy from Eq. (\ref{eq17}).
It turned out that an accurate value of $A'_{\rm I}$ can be obtained by solving
the equation (\ref{eq21}) for $A'_{\rm I}$ at the value of the eigenvalue which 
had been determined before.

It is found that there are 3 classes of eigenfunctions: ''external'' or 
''class 1''-states which are predominantly localized in the vacuum outside
of the insulator $(r > R_1)$, ''internal'' or ''class 2''-states which are mainly
localized within the metallic core ($r<R_2$), and ''class 3''-states which 
exhibit a non-negligible probability amplitude both in region 0 and region II. 
For a sufficiently large value of $B$, the probability for finding the electron
in region I, i.e. within the insulator, turns out to be negligibly small.
The spectrum of class 1-states resembles the Coulomb spectrum. On the other 
hand, the eigenvalues of class 2-states, in lowest approximation, are given by
the eigenvalues of an infinite spherical square well of radius $R_2$
i.e. by the boundary condition $j_l(\kappa_{\rm II}R_2) = 0$.

If $\kappa_{\rm II}R_2 \gg 1$, the spherical Bessel function $j_l(\kappa_{\rm II}
R_2)$ can be approximated by
\begin{equation}
 j_l(\kappa_{\rm II}R_2) \approx {1\over\kappa_{\rm II}R_2} \cos\left[\kappa_{\rm II}R_2 - 
 {l\pi\over 2} - {\pi\over 4}\right] \,\,.
\label{eq28}
\end{equation}
The zeros of this function together with (\ref{eq2.11}) lead to the energy 
formula
\begin{equation}
 E_{nl} = = -V_{\rm II} - {Ze^2_0 \over R_2} + {\hbar^2\pi^2\over 2MR^2_2} 
 \left({3\over 4} + n + {l\over 2}\right)^2\,,
\label{eq29}
\end{equation}
where $n = 0,1,\ldots$ as well as $l$.

The spacing between neighbouring levels of given $l$ or given $n$ is
\begin{eqnarray}
&& E_{n+1,l} - E_{nl} = {\hbar^2\pi^2\over 2MR^2_2} \left({5\over 2} 
   + 2n + 2l\right) \,\,,
\label{eq30} \\
&& E_{n,l+1} - E_{n,l} = {\hbar^2\pi^2\over 2MR^2_2} \left(1 + n 
   + {l\over 2}\right)\,\,.
\label{eq31}
\end{eqnarray}
One should notice the completely different spacing of the eigenvalues of 
internally located electrons from the Coulomb dominated spacing of 
externally located electrons. Of course, if $\kappa_{\rm II}R_2$ is not 
large compared to 1, the simple results (\ref{eq28}) to (\ref{eq31})
do not hold.

If it happens that, for given $l$, the energy of a predominantly externally
and of a predominantly internally localized state come very close to each other,
the corresponding eigenfunctions are not negligible in both the regions II and 
0. We call eigenstates of this nature ''class 3''-states. These states play an 
important role in the lifetime of the system and we shall come back to them in 
Subsec. 4.2.

\begin{figure}
\label{fig3}
\begin{center}
\includegraphics[width=10cm]{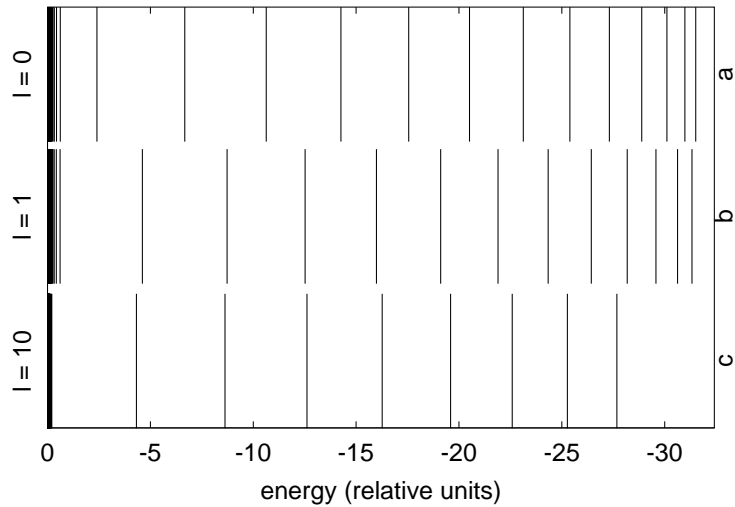}
\caption{Comparison of the spectra of $l = 0$ states (Fig.~3a) and $l=1$ 
states (Fig.~3b) and $l=10$ states (Fig. 3a) for the following set of 
parameters: $Z = 1$, $R_1 = 28$~\AA; $R_2 = 20$~\AA; $V_{\rm II} = 15950$~meV;
$b = 5.83$; $v_{\rm II} = 31.031$, $R_0 = 0.2 71$~\AA; $\varepsilon_{\rm
min}=0.1$; $\varepsilon_{\rm max}=32.41$; energy mesh $\Delta \varepsilon =
0.0001$. Energy unit: $Z e^2_0/R_1=514$ meV.}
\end{center}
\end{figure}

In Fig.~3, we show the spectrum of eigenvalues $\varepsilon$ for $l=0$, $l=1$,
and $l=10$ for a given set of parameters. We note that, apart from the
dimensionless parameters defined in the text, lengths are given in 
{\AA}ngstr\"om (10$^{-10}$~m) and energies in meV (10$^{-3}$~eV).

Given the energy unit $Ze^2_0/R_1 = 514$~meV in Fig.~3, the 
$|\varepsilon|$-values belonging to externally located states must be smaller 
than 514~meV. The lowest-lying external state corresponds to the lowest line of
the bunch of narrowly spaced lines at the top of the  spectrum at $\varepsilon
= 0,68$ or $E = -350$~meV. Note that the energy  difference between the first
excited external $s$-state and the lowest  external $s$-state is about
69,5~meV. This is also the order of magnitude of the average distance between
low-lying external levels. Since the ''room temperature'' of 300$^\circ$~K
corresponds to 26~meV, one may conclude  that, at room temperature, for the
parameter set chosen in Fig.~3, the  externally bound electrons are
predominantly in their lowest available states.

The spectrum of eigenvalues looks similarly for different fixed values of the
angular momentum $l$. The fact that we show only the subsets of eigenvalues for
given $l$-values in Fig.~3 should not lead to the erroneous impression that the
levels of internally located electrons are so widely  spaced that in general,
no level of internal electrons is to be expected in the range of energies of
external electrons. This is not so due to the fact that there are also the
levels for $l \neq 0$. Consequently, there are generally ''intruders'' of
internal levels in the spectrum of external ones and, consequently, there is in
general also a limited number of ''class  3-states'' which. as we shall show,
lead to rapid decay.

We should also notice in Fig.~3 that the distance between successive
eigenvalues for given $l$ increases with increasing eigenvalue, i.e. as 
$|\varepsilon|\rightarrow 0$. The reason for this trend is qualitatively seen
from the formulae (\ref{eq30}) and (\ref{eq31}). For the parameter set chosen
in Fig.~3, these approximate formulae are only valid for the high-lying levels.

\begin{figure}
\label{fig4}
\begin{center}
\includegraphics[width=10cm]{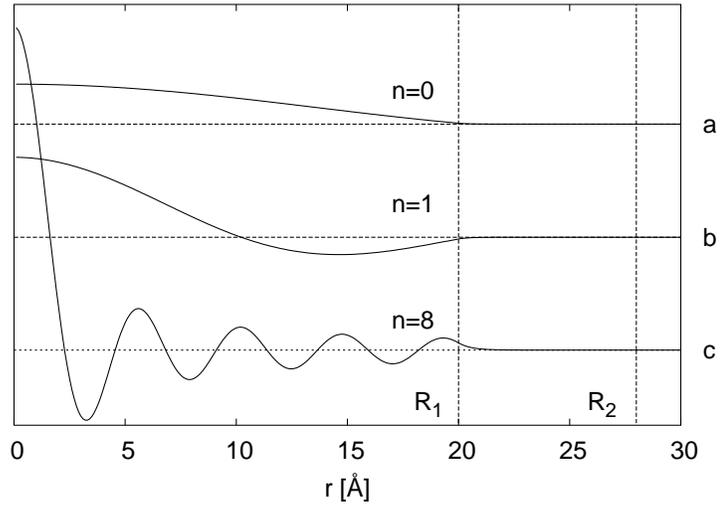}
\caption{Internally located wavefunction (class 2-states) with radial node
number $n = 0$ (Fig.~4a), $n = 1$ (Fig.~4b), and $n = 8$ (Fig.~4c). Orbital
angular momentum $l = 0$. Energies (in units of $Ze^2_0/R_1 = 514.3~meV$):
$|\varepsilon(n=0; l = 0)| = 32,235$; $|\varepsilon(n=1; l = 0)| = 31,698$; 
$|\varepsilon(n=8; l = 0)| = 17.963$. Choice of parameters as in Fig.~3.}
\end{center}
\end{figure}

In Fig.~4, we show the wavefunctions corresponding to internal $s$-states with
0, 1, and 8 radial nodes. It should be noted that even for the case of the
highest one with 8 radial nodes at $|\varepsilon| = 17,96$, the  localization
within region II is perfect and the tails in region I are very  tiny.

\begin{figure}
\label{fig5}
\begin{center}
\includegraphics[width=10cm]{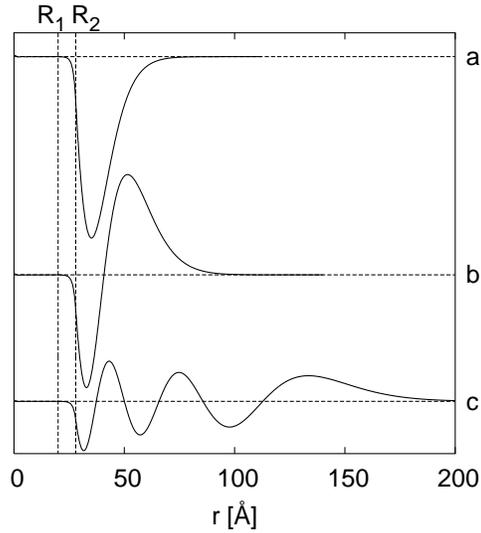}
\caption{Externally located wavefunctions (class 1-states) with radial
node number $n = 0$ (Fig.~5a), $n = 1$ (Fig.~5b) and $n = 5$ (Fig.~5c).
Orbital angular momentum $l = 0$ energies (in units of $Ze^2_0/R_1
= 514.3$~meV): $|\varepsilon (n = 0; l = 0)| = 0.63325$;
$|\varepsilon(n = 1; l = 0)| = 0.44533$; $|\varepsilon (n = 5; l = 0)| =
0.18213$. Choice of parameters as in Fig.~3.}
\end{center}
\end{figure}

In Fig.~5, we present a comparison of external $s$-wavefunctions for radial
node numbers $n = 0,1$, and 5. Again one notices that the wavefunctions dive
only weakly into region I and are completely negligible in region II. The
energy difference ($E_{n = 1,l=0} - E_{n=0,l=0}$) between the two lowest-lying 
external levels turns out to be 96~meV. The parameter set was chosen to be the
same for the Figs.~3--5.

For the high-lying eigenenergy $E_{n=8,l=0} \approx -93,5$~meV the most
external maximum of the wave function is seen to occur at $r \approx 130$~\AA ~
whereas the wavefunction of the lowest external $s$-state with $n=0$ becomes
vanishingly small already at $r \approx 60$\AA. This strong dependence of the
radial extension on the energy of the state is typical for a Coulomb-spectrum.

The class 1- and class 2-states are thus of very different nature. The 
eigenvalues of class 1-states must be above the potential energy 
$\left(-Ze^2_0/R_1\right)$ on the surface $S_1$ of the insulator. Their spacing
is much smaller than the one of class 2-states and tends to zero  as the
binding energy approaches zero. The lowering of the well-depth $-V^0_{\rm II}$
by the potential produced by the positive surface charge on the metal surface
$S_2$ and the screening effects in the insulator is given by
$-\left[Ze^2_0/(\varepsilon_{\rm I} R_2) + Ze^2_0(\varepsilon_I -
1)/(\varepsilon_{\rm I}R_1)\right]$. This lowering of the Fermi energy $E_F$ in
the metal is larger than the minimal binding energy $-\left[Ze^2_0/R_1\right]$
of the lowest external electron by the amount
\begin{equation}
 \Delta E: = -{Ze^2_0\over\varepsilon_{\rm I}} \left[{1\over R_2} + {\varepsilon_{\rm I} - 1
\over R_1}\right] + {Ze^2_0\over R_1} = -{Ze^2_0\over \varepsilon_{\rm I}}
 \left({1\over R_2} - {1\over R_1}\right)
\label{eq32}
\end{equation}
For the choice $R_1=28$~\AA ~and $R_2=20$~\AA ~and the charge number $Z=1$,
we have
$$
 \Delta E = -{205,7~{\rm meV}\over\varepsilon_{\rm I}}
$$
i.e. a number between --20 and --200~meV depending on the dielectric constant
$\varepsilon_{\rm I}$ of the insulator.

If there are unoccupied levels between the Fermi energy $E_F$ and $(E_F + 
|\Delta E|)$, an electron in the lowest external state can decay into a
lower-lying  internal state.


\subsection{Electromagnetic decay of external states}

For the sake of simplicity, let us consider a system with one external electron
and $Z$ positive surplus charges. If the external electron is in some arbitrary
externally located state it will first decay to the lowest-lying external state
by a cascade of unhindered transitions. This lowest-lying external state is
characterized by the angular momentum $l=0$ and has no radial nodes ($n=0$).

Subsequently, this lowest class 1-state will decay into lower-lying unoccupied
internally located states (class 2-states). The strongest transitions are of
multipolarity E1 and we only consider such dipole transitions. Even these most
rapid transitions between a class 1 and a class 2 state are reduced in comparison
with transitions between states of the same class due to the small overlap 
between externally and internally located wavefunctions.

The probability per time-unit for a dipole transition from the initial state
$\psi_i$ to the final state $\psi_f$ is given by \cite{Brans}
\begin{equation}
 W_{12} = {4\alpha c\over 3} {|E_1 - E_2|^3\over (\hbar c)^3} |\langle\psi_1|
 \vec r|\psi_2\rangle|^2  \,\,,
\label{eq33}
\end{equation}
where $E_2$ and $E_1$ denote the final and initial energy, resp., and 
$\alpha \equiv e^2_0/(\hbar c)$ is the fine structure constant.

The initial and final wavefunctions are
\begin{eqnarray}
&& \psi_1(r,\vartheta,\varphi) =  \tilde R_{l_1}(r,E_1) Y_{l_1m_1} (\vartheta,
 \varphi) \,\,,
\label{eq34} \\
&& \psi_2(r,\vartheta\varphi) = \tilde R_{l_2}(r,E_2) Y_{l_2m_2} (\vartheta,
 \varphi) \,\,,
\label{eq35}
\end{eqnarray}
where we leave away the spin-functions, and where $\tilde R_{l_1}$, $\tilde 
R_{l_2}$ are the properly normalized radial functions (\ref{eq2.43}). 
The matrix-element factorizes
\begin{equation}
\langle\psi_2|\vec r|\psi_1\rangle = \int\limits^\infty_0 dr \tilde R_{l_2}(r)
 r^3\tilde R_{l_1}(r)\cdot\int d\Omega \, Y^\ast_{l_2m_2} \vec e_r \,Y_{l_1m_1}
\,\,. \label{eq36}
\end{equation}
It is easily seen that the following relation holds:
\begin{equation} \left[\int d\Omega Y^\ast_{l_1m_1} \vec e_r Y_{l_2m_2}\right]^2 = 
 |W_+|^2 + |W_-|^2 + |W_z|^2\,,
\label{eq37}
\end{equation}
where the quantities $W_{\pm}$ and $W_z$ are defined by
\begin{eqnarray}
&& W_{\pm} : =  \mp\sqrt{{4\pi\over 3}} \int d\Omega Y^\ast_{l_1m_1}
 Y_{1,\pm 1}Y_{l_2m_2} \,\,,
\label{eq38} \\
&& W_z: = \sqrt{{4\pi\over 3}} \int d\Omega Y^\ast_{l_1m_1} Y_{10} Y_{l_2m_2}
\,\,.
\label{eq39}
\end{eqnarray}
They are evaluated in terms of Clebsch-Gordan coefficients using the Wigner-Eckart
theorem \cite{Rose}: 
\begin{eqnarray}
&& W_{\pm} : =  \mp\sqrt{{2l_2 +1\over 2l_1+1}} C(l_2 1l_1,m_2, \pm 1, m_1)
 C(l_2 1 l_1 000)  \,\,,
\label{eq40} \\
&& W_z: = \sqrt{{2l_2 +1\over 2l_1+1}} C(l_2 1l_1m_2 0 m_1)
 C(l_2 1 l_1 000) \,\,.
\label{eq41}
\end{eqnarray}
The non-vanishing values for $W_{\pm}$ and $W_z$ are presented in Table~2.
The radial integral, i.e. the 1$^{\rm st}$ factor in Eq.~(\ref{eq36}) is
evaluated numerically.

\begin{table}
\caption{The quantities $W_z$ and $W_{\pm}$ entering the dipole transition
probability (see Eqs. (\ref{eq38}), (\ref{eq39})).
\label{tab2}}
\begin{center}
\begin{tabular}{|l|c|c|}
\hline
 &$l_1 - l_2 = 1$   & $l_1 - l_2 = -1$ \\
\hline
 $m_2 = m_1 + 1$ &
 $W_- = \left[{(l_1 - m_1 - 1)(l_1 - m_1)\over (2l_1 + 1)(2l_1 - 1)\cdot 2}
        \right]^{{1\over 2}}$ &
 $W_- = -\left[{(l_1 + m_1 + 1)(l_1 + m_1 + 2)\over (2l_1 + 1)(2l_1 + 3)\cdot 2}
         \right]^{{1\over 2}}$ \\
\hline
 $m_2 = m_1$ &
 $W_z = \left[{(l_1 - m_1)(l_1 + m_1)\over (2l_1 + 1)(2l_1 - 1)}
        \right]^{{1\over 2}}$ &
 $W_z = \left[{(l_1 - m_1 + 1)(l_1 - m_1 + 2)\over (2l_1 + 1)(2l_1 + 3)}
        \right]^{{1\over 2}}$ \\
\hline
 $m_2 = m_1 - 1$ &
 $W_+ =-\left[{(l_1 + m_1)(l_1 + m_1 + 1)\over (2l_1 + 1)(2l_1 - 1)\cdot 2}
        \right]^{{1\over 2}}$  &
 $W_+ = \left[{(l_1 - m_1 + 1)(l_1 - m_1 + 2)\over (2l_1 + 1)(2l_1 + 3)\cdot 2}
        \right]^{{1\over 2}}$  \\
\hline
\end{tabular}
\end{center}
\end{table}

We consider the case that the initial state is the lowest external state,
i.e. a nodeless wavefunction with $l_1 = 0$. Thus, the final angular
momentum $l_2$ is equal to 1.

The final magnetic moments $m_2$ may have the value $m_2 = -1, 0, +1$.

We are mainly interested in the lifetime $\tau$ of a ''mesoscopic atom'' which
is the inverse of the total decay rate $P$:
\begin{equation}
 \tau: = {1\over P}\,\,.
\label{eq42}
\end{equation}
The total decay rate $P$ is defined by summing the transition probabilities
$W_{12}$ over all the unoccupied final states $\psi_2$ which are connected with
the initial state $\psi_1$ by a finite matrix element $\langle\psi_1|\vec r|
\psi_2\rangle$:
\begin{equation}
 P = {\sum_2}'W_{12} \,\,.
\label{eq43}
\end{equation}
Considering only dipole transitions is, of course, an approximation which
yields only a lower limit for the total transition rate and an upper limit
for the lifetime.

The restriction of the sum in (\ref{eq43}) to unoccupied internal electronic
states is indicated by the prime '.

At temperature 0, a dipole transition starting from the initial external 
$s$-state can go into all the $l_2 = 1$ states the energy of which is between
the Fermi energy $E^0_F$ and the energy $E_1$ of the initial state
$$
 E^0_F < E_2 < E_1 \,\,.
$$
At finite temperature, one has to weigh all the final states with the
factor
$$
 \left[1 - {1\over \exp\left({E_2 - \mu\over T}\right) + 1}\right]\,\,.
$$
So far, we calculated the lifetime for the case $T = 0$ only.

\begin{table}
\caption{Separation energy $W_0$ of an electron, Fermi-energy $E^0_F$ and
well depth $V^0_{\rm II}$ of uncharged metals (in meV).
\label{tab3}}
\begin{center}
\begin{tabular}{|l|c|c|c|c|c|c|c|}
\hline
 & Be & Na & Al & Ca & Fe & Cs & Au \\
\hline
$W_0$ & 3920 & 2350 & 4250 & 2800 & 4310 & 1810 & 4300 \\
\hline
$E^0_F$ & 14300 & 3240 & 11700 & 4690 & 11100 & 1590 & 5530 \\
\hline
$V^0_{\rm II}$ & 18220 & 5590 & 15950 & 7490 & 15410 & 3400 & 9830 \\
\hline
\end{tabular}
\end{center}
\end{table}

In Table~3, we present the well depth $V^0_{\rm II}$, the separation energy
$W_0$, and the Fermi energy $E^0_F = V^0_{\rm II} - W_0$ for a couple of metals.
We note that these empirical parameters have been determined for materials with
macroscopic size, not for mesoscopic clusters of such materials. Nevertheless,
the parameters can be expected to be essentially correct also for mesoscopic
systems.

Let us now come to the results we obtained for the lifetime as a function
of the different parameters of the system. We studied separately the dependence
of the lifetime $\tau$ on the parameters $R_1$, $B$, $Z$, and $R_2$, keeping 
always all but one parameter fixed.

\bigskip
\noindent
\underline{Dependence of $\tau$ on $R_1$}

\medskip
\begin{table}
\caption{Lifetime $\tau$ as a function of the outer radius $R_1$~(\AA)
for: $Z = 1$; $R_2 = 20$~\AA; $B = 3000$~meV; $V^0_{\rm II} = 15950$~meV (Al).
Lifetimes in units: sec, minute (min), day (d), year (a).
\label{tab4}}
\begin{center}
\begin{tabular}{|l|c|c|c|c|c|c|c|c|}
\hline
$R_1$ & 26 & 28 & 30 & 32 & 34 & 36 & 38 & 40 \\
\hline
$\tau$ & 4.83 sec & 52.9 sec & 20.4 min & 9.76 h & 12.9 d & 432 d & 40.8 a & 1431 a \\
\hline
\end{tabular}
\end{center}
\end{table}

Augmenting $R_1$ for fixed value of $R_2$ enlarges the width of the barrier
and, consequently, decreases the amplitude which an externally located electron
has in the inner region II. As the decay of a class 1-state into a class-2
state depends strongly on this tail, the transition probability must decrease
and the lifetime must increase as a function of rising $R_1$. The results 
presented in Table~4 support this expectation. The lifetime $\tau$ is found 
to rise roughly exponentially with $(R_1 - R_2)$. This is so because the 
amplitude of the tail in region II depends exponentially on $(R_1 - R_2)$
as one can see from WKB. The dependence of $\tau$ on $R_1$ is also shown
in Fig.~6 in logarithmic scale.

\begin{figure}
\label{fig6}
\begin{center}
\includegraphics[width=7cm]{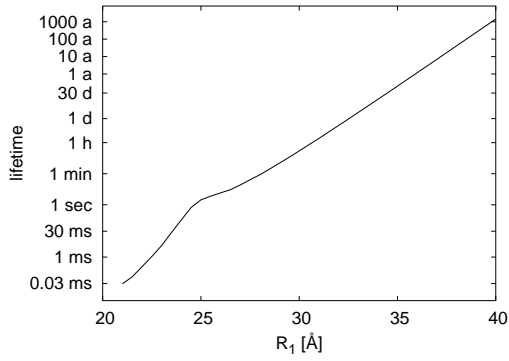}
\caption{Lifetime in sec and logarithmic scale as a function of
radius $R_1$. Apart from $R_1$ , all the other parameters are chosen as
in Fig.~3.}
\end{center}
\end{figure}

\bigskip
\noindent
\underline{Dependence of $\tau$ on the barrier height $B$}

\medskip
In Fig.~7, we show the dependence of the lifetime $\tau$ on the barrier height
$B$ in logarithmic scale. The lifetime increases as a function of rising $B$.
The increase is somewhat smaller than exponential as can be expected from
WKB.

\begin{figure}
\label{fig7}
\begin{center}
\includegraphics[width=7cm]{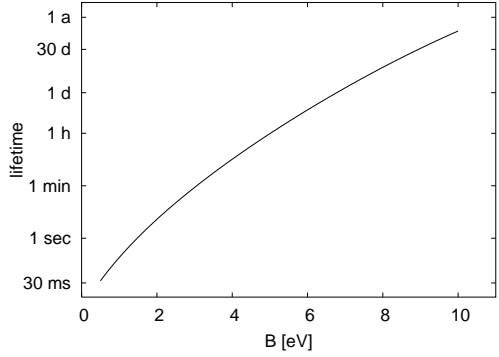}
\caption{Lifetime $\tau$ in sec and logarithmic scale as a function of
the barrier height $B$. Apart from $B$ , all the other parameters are chosen as
in Fig.~3.}
\end{center}
\end{figure}

\bigskip
\noindent
\underline{Dependence of $\tau$ on $R_2$ and on $Z$}

\medskip

\begin{figure}
\label{fig8}
\begin{center}
\includegraphics[width=7cm]{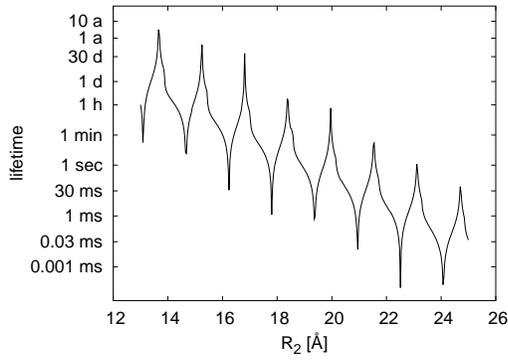}
\caption{Lifetime $\tau$ in sec and logarithmic scale as a function of
$R_2$. Apart from $R_2$ , all the other parameters are chosen as in Fig.~3.}
\end{center}
\end{figure}

If one changes $R_2$ with all the other parameters fixed, one modifies predominantly
the internally located class 2 states and the corresponding energies.
As one can see from Fig.~8, the lifetime $\tau$ as a function of rising
$R_2$ decreases on the average as one would expect from the fact that the 
thickness $(R_1 - R_2)$ of the surface layer decreases. However, beside the 
average decrease of the lifetime, one observes a kind of resonance structure:
Within a change of $R_2$ by 1~\AA, the lifetime $\tau$ can change by 10 orders
of magnitude.\\
How can we understand these huge fluctuations of the lifetime?

\begin{figure}
\label{fig9}
\begin{center}
\includegraphics[width=7cm]{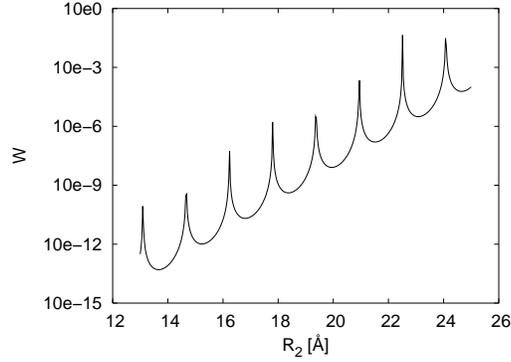}
\caption{Probability ($W$) of lowest external electron to be found in the
inner region II ($R < R_2$) as a function of $R_2$.
Apart from $R_2$ , all the other parameters are chosen as
in Fig.~3.}
\end{center}
\end{figure}

In Fig.~9, we show the probability of an externally located electron to be
found in the inner region II. This localization probability in region II, which
is of crucial importance for the transition rate, is seen to show pronounced 
peaks at precisely those values of $R_2$, where the lifetime has very small
values as one can see by comparing Fig.~8 and Fig.~9. The large variations of
the transition rate as a function of $R_2$ (and of $Z$) can be simply explained
as follows:\\ As one changes $R_2$ (or $Z$), the eigenvalues of class 2-states
move whereas the ones of class 1-states remain essentially the same. More
specifically, the eigenvalues $E_{nl} = -|E_{nl}|$ of class-2 states are
lowered, if the inner radius $R_2$ or the positive surplus charge $Z$ is
increased. Thereby, it happens regularly that a class 2-state which has the
same angular momentum $l$ as the decaying class 1-state, crosses the energy of
this class 1-state. If one augments $R_2$ or $Z$ on a sufficiently long scale,
such crossings occur several times. Close to the crossing, the external class
1-state develops a non-negligible amplitude within region II and the class
2-state develops an  appreciable amplitude in the external region 0. We call
such mixed states ''class 3-states''. If the decaying class 1-state happens to
have the nature of a class 3-state, its decay probability to lower-lying class
2-states is considerably enhanced as compared to the ordinary situation where
no class 2-state of the same angular momentum is available. We note in passing
that this enhancement of the transition probability to lower-lying class
2-states does not only occur for dipole transitions but also for transitions of
higher multipolarity. So far we have not yet calculated the transition rates of
higher multipolarity, because they would only matter if the decaying external
state is a class 3-state.

The phenomenon can be considered as an example of the mechanism which Landau
and Zener had studied in famous papers \cite{Lan32}. This is demonstrated in 
Appendix.

\begin{figure}
\label{fig10}
\begin{center}
\includegraphics[width=10cm, angle=90]{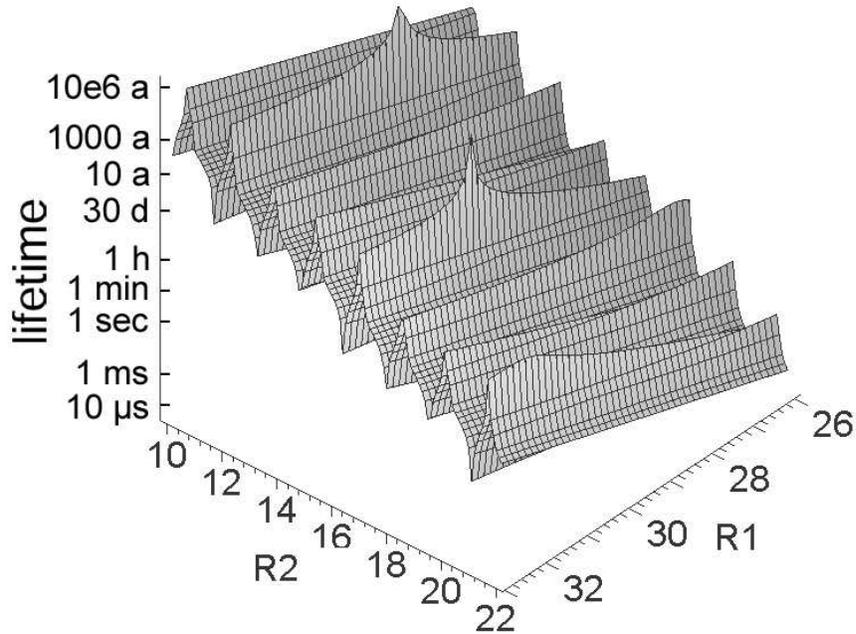}
\caption{Lifetime  $\tau$ (sec, log.scale) as a function of radius $R_1$ and 
$R_2$ (in \AA) in a 3-dimensional plot. Except for $R_1$ and $R_2$, all the 
parameters are chosen as in Fig.~3.}
\end{center}
\end{figure}

In Fig.~10, the dependence of the lifetime $\tau$ on the radii $R_1$ and $R_2$
is shown in a 3-dimensional plot. Again one notices that the dependence of 
$\tau$ on $R_1$ is smooth whereas the one on $R_2$ contains ''resonances''.

\begin{figure}
\label{fig11}
\begin{center}
\includegraphics[width=10cm]{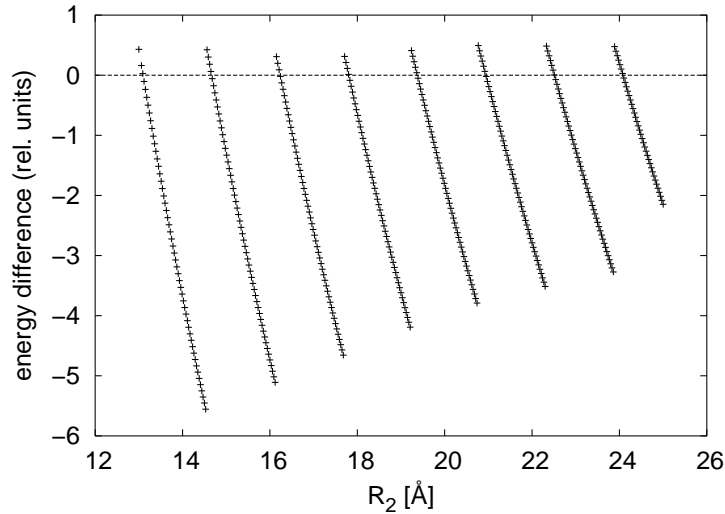}
\caption{Energy difference between lowest-lying class 1-state and the closest 
lying class 2-state of the same angular momentum as a function of $R_2$ (\AA).
Energies are given in units of $Ze^2_0/R_1 = 514.3$~meV. Except for 
$R_2$, all the other parameters are chosen as in Fig.~3.}
\end{center}
\end{figure}

In Fig.~11, the energy difference between the lowest-lying class 1 state and 
the closest lying class 2-state of the same angular momentum is shown. The plot
supports the explanation of the resonance phenomenon in terms of a coupling
between close-lying states.

In the numerical results presented in this section, the values $B$ of the barrier
in region I and the well depth $V_{\rm II}$ of the average potential in region II
and the dielectric constant $\varepsilon_{\rm I}$ were chosen arbitrarily, and not
determined for given choices of the material.


\section{Summary and Discussion}
\label{sec:6}

In this paper, we determined the quantum states and
eigenenergies of an electron in the average potential produced
by a positively charged metal cluster which is covered by an
insulating surface layer. Because of the similarity of the
externally localized electronic states with the bound states of
an electron in the Coulomb potential of an atom, we occasionally
referred to the system as a ''mesoscopic atom'' or, in case of a
non-zero total charge, as a ''mesoscopic ion''.

In particular, we investigated the electromagnetic decay of an
externally located electron (''class 1-state'') to a lower-lying
unoccupied electronic state located within the metal (''class
2-state''). It turned out that the life-time depends sensitively
on the radius $R_2$ of the metal core and on the number $Z$ of
positive surplus charges of the metal. In the model we studied,
we found life-times varying between 10$^{-10}$~s and years
depending on the value of $R_2$ or of $Z$.

The value $B$ of the barrier height was 3 eV in our calculations.
The lifetime depends strongly on the value of this parameter and
decreases for decreasing $B$.

The origin of the large fluctuations was found to be the
appearance of states with non-negligible amplitudes both
outside \underline{and} inside of the insulating surface layer
(''class 3-states'').  The mechanism leading to these ''class
3-states'' is described in Appendix.

The wavefunctions and eigenenergies were calculated neglecting
the polarization of the insulating surface layer by externally
located electrons. The potential produced by these electronic
polarization effects was determined in Sec.~3. It was shown
that this additional potential matters mainly in the vicinity
of the outer surface $S_1$ of the insulator and decreases
exponentially outside of this surface.  The polarization
potential could be taken into account by diagonalizing the total
Hamiltonian including the polarization potential in the basis of
the wavefunctions determined in this paper.

Apart from this physical simplification of the problem we
neglected the Coulomb potential produced by the positive surplus
charge within the insulator. This approximation is justified
whenever the change of the Coulomb potential within the
insulating surface layer is much smaller than the barrier height
$B_0$. This condition was fulfilled in all the cases we studied
numerically.

Let us now turn to the discussion of some open questions:

\medskip
\noindent
\underline{Is the model sufficiently realistic?}\\
Certainly, our model should only be regarded as a first step which
has to be improved in various respects.

The 1$^{\rm st}$ improvement we already mentioned is to
incorporate the additional potential between the external
electron and the polarization cloud it produces in the
insulator. The form of this electronic polarization potential
was derived in Sec.~3.

A 2$^{\rm nd}$ improvement would be to replace the constant
potentials $-V^0_{\rm II}$ in the metal and $+B_0$ in the
insulator by periodic potentials defined by the crystal
structure of these materials. Of course, this is much more
complicated and, to our knowledge, it has rarely been done so
far for mesoscopic systems. Furthermore, the crystal structure in mesoscopic
aggregates is in general poorly known.

As a result, the electronic eigenvalues of the total system are
expected to exhibit a band structure similarly to a macroscopic
solid, the insulating surface layer being characterized by a
completely filled valence band and an empty conduction band. The
empty conduction band in the insulator can be at negative or at
positive energy if the energy of the neutral system plus one
electron at infinity is put equal to 0.  Describing the average
potential in the insulator by a positive barrier $B_0 > 0$ is
only meaningful if the conduction band of the insulator is at
positive energy.

As a further restrictive condition for our model to be a reasonable 1$^{\rm
st}$ approximation there should be no localized traps for an additional
electron in the uncharged insulator. This implies that the individual atoms or
molecules of the insulator should not have bound states with an additional
electron as a monomer \cite{Haber}. In general, bound states of an  electron
with a neutral atom or molecule do not exist for atoms or molecules with
completely filled shells. Therefore, one should use insulators which exhibit
this property. Presumably, Al$_2$0$_3$ fulfills this condition. Finally, we
neglected effects of the electron-phonon coupling altogether. We  presume that
they tend to reduce the lifetime, but perhaps not drastically.

\bigskip
\noindent
\underline{What were the physical motivations of this investigation}

\medskip
As long as one can only produce a limited number of ''mesoscopic
atoms'', one could just investigate experimentally the effects
which we studied in the present paper, especially the amusing
strong dependence of the lifetime on the radius and the charge
of the metallic core. Although we would be highly pleased by an
experimental scrutiny of our predictions, still more exciting
open questions could be asked, if it were possible to produce
longlived systems of the kind we presented in this paper, and in
large quantities. In this case, one would have access to the
study of molecules formed from mesoscopic atoms and to condensed
matter consisting of mesoscopic atoms as building blocks.

The presence of externally localized electrons would represent a
fundamental difference from the case of a condensed system of
uncharged mesoscopic clusters. On the one hand, one may expect
that, in a condensed matter phase of mesoscopic atoms, the
weakly bound external electrons would partly dissociate from
their mother atoms and move freely in the average field produced
by the charged insulated cores. We would thus surmise that such a
system would be a (good) conductor.  On the other hand, the fact that
the building blocks of such a condensed matter carry charge,
will have a profound influence on the equilibrium state of the
material. Probably, its density is lower than for uncharged
constituents, where the cohesive forces are of van der Waals type
and, consequently, of much shorter range.

\bigskip
\noindent
\underline{How far are such dreams from reality?}\\
There are evidently two fundamental obstacles to overcome, a theoretical
one and an experimental one.

The theoretical problem is the stability of the charged and
insulated single system, and the experimental one is the
production of a large number (10$^{23}$) of such systems.

As we learnt from this paper, a mesoscopic system of the kind we
investigated can only be stable in a strict sense, if there is
no empty internally located electronic state below the energy of
the lowest externally located state which is at an energy
slightly above $-Ze^2_0/R_1$.

The positive surplus charge $Ze_0$ on the surface $S_2$ of the
metal leads to a decrease of the well-depth $V^0_{\rm
II}\rightarrow V_{\rm II}$ as we have shown in Sec.~3:
\begin{equation}
  V_{\rm II} = V^0_{\rm II} - {Ze^2_0\over\varepsilon_{\rm I}}
  \left({1\over R_2} + {\varepsilon_{\rm I} - 1\over R_1}\right) \,\,.
\label{eqS1}
\end{equation}
 If, for the uncharged metal, the Fermi level is at the energy 
$E^0_F - V^0_{\rm II}$ (with $V^0_{\rm II}$, $E^0_F > 0$), it
lies at the modified energy
\begin{equation}
 E^0_F - V_{\rm II} = E^0_F - V^0_{\rm II} + {Ze^2_0\over\varepsilon_{\rm I}}
  \left({1\over R_2} + {\varepsilon_{\rm I} - 1\over R_1}\right)
\label{eqS2}
\end{equation}
in the charged metal.

So the condition for strict stability is that there is no level of an
internally located electron between $E^0_F - V_{\rm II} < 0$ and 
$-Ze^2_0/R_1$. As the separation energy $W_0 > 0$ of an
electron from the uncharged metal and the Fermi energy $E^0_F$ are 
related to $V^0_{\rm II}$ by
\begin{equation}
 V^0_{\rm II} = E^0_F + W_0\,,
\label{eqS3}
\end{equation}
we can state that strict stability is achieved whenever there is no level
between the energies $-\left\{W_0 - Ze^2_0/\varepsilon_{\rm I}\cdot
\left[1/R_2 + (\varepsilon_{\rm I} - 1)/R_1\right]\right\}$
and $-Ze^2_0/R_1$.
This means that, metal with small $W_0$ is favourable for stability.

In an ordinary metal, the separation energy $W_0$ is of the order of a couple
of eV (examples: $W_0(\rm Al) = 4,28$~eV; $W_0(\rm Ag) = 4,26$~eV).
In the case of a radius $R_2 = 20$~\AA ~of the metal core, the distance 
between neighbouring electronic levels is of the order of 100~meV. This means
that in the case we chose for our numerical investigations, there are many
unoccupied levels below the energy of the lowest external electron. However,
if we consider a system with an inner radius $R_2 \approx 4$~\AA, the spacing
of internally located electron levels of given $l$ approaches already 1~eV.
It is therefore conceivable that one might succeed to model a strictly stable
,,mesoscopic atom''.

A further trick to widen the spectrum of internally localized electrons
would be to replace the metallic sphere with radius $R_2$ by a metallic bubble
of outer radius $R_2$ and a thickness of 1~\AA~to 2~\AA.~In this case the levels
differing by the number of radial nodes would have a much larger spacing.

\medskip
\noindent
\underline{Can such a system be made experimentally?}

This is perhaps possible, if one replaces the charged metal
layer by a (charged) fullerene $C_{60}$, which has a radius of
about 3,5~\AA. The separation energy $W_0$ of an electron from
graphite is known to be $W_0 = 3,8$~eV. We surmise that the
separation energy from the fullerene is about the same. Of
course, it would no longer be justified to describe free
electrons in the fullerene as moving in a simple square well
perpendicular to the sphere $S_2$. But the spacing of electronic
levels is expected to be in the order of eV.

Finally, one is not limited to the spherical geometry. We may also envisage 
to consider charged nanotubes consisting of a cylindrical surface crystal 
of $C$-atoms covered by an insulating surface layer. The charge could in 
this case be generated by applying an external voltage to the ends of the
nanotubes. As there is a lot of activity spent on nanotubes presently, we hope
that one could succeed to produce small macroscopic amounts of insulated 
charged nanotubes in a not too far future. Of course, the problem of the 
stability of externally localized electrons with respect to electromagnetic
decay would be similar for nanotubes as for the case of the system we 
investigated. So we hope that our investigation stimulates the interest of
the experts in the field of mesoscopic physics.


\bigskip

\vspace{1cm}

\noindent
{\bf Acknowledgement:}\\
Klaus Dietrich gratefully acknowledges the hospitality and support on the
occasion of his stay at the Theoretical Physics Institute of the UMCS
in April 2003. Krzysztof Pomorski is indebted to the Physics Department of the
Technical University of Munich for granting of the honorarium of a visiting 
professor position in February 2002.

\newpage

\centerline{\Large\bf Appendix}

\bigskip
\begin{center}
{\bf Landau-Zener Model for explaining the appearance of class 3-states}
\end{center}

Let us consider the simplified potential
\begin{equation}
 \widehat V(r) =  -V_{\rm II} \theta_0(R_2 - r) + B\theta_0(R_1 - r)
 \theta_0(r - R_2) - {Ze^2_0\over r}\theta_0 (r - R_1) \,\,.
\label{eqA1}
\end{equation}
We decompose this potential as follows
\begin{equation}
\widehat V(r) = \tilde V_1(r) + \tilde V_2(r) \,\,,
\label{eqA2}
\end{equation}
%
\begin{equation}
 \widehat V_1(r) = \theta_0(r - R_1) \left[-B - {Ze^2_0\over r}\right] \,\,,
\label{eqA3}
\end{equation}
%
\begin{equation}
 \widehat V_2(r) = -V_{\rm II}\theta_0 (R_2 - r) + B\theta_0(r - R_2)
\label{eqA4}
\end{equation}
and define the two Hamiltonians
\begin{equation}
 \widehat H_{1(2)}: = \widehat T + \widehat V_{1(2)}(r)\,\,,
\label{eqA5}
\end{equation}
where $\widehat T$ is the kinetic energy operator of radial motion
\begin{equation}
 \widehat T = -{\hbar^2\over 2M} \left({d^2\over dr^2} + {2\over r}{d\over dr}
 - {l(l +1)\over r^2}\right)\,\,.
\label{eqA6}
\end{equation}
The eigenstates $\varphi^{(1)}_\nu$, $\varphi^{(2)}_\nu$ of the Hamiltonians
(\ref{eqA5})
\begin{equation}
 \widehat H_1 \varphi_\nu^{(1)}(r) = E^{(1)}_\nu \varphi^{(1)}_\nu \,\,,
\label{eqA7}
\end{equation}
%
\begin{equation}
  \widehat H_2 \varphi_\nu^{(2)}(r) = E^{(2)}_\nu \varphi^{(2)}_\nu
\label{eqA8}
\end{equation}
form a non-orthogonal set of basis functions.

We consider an eigenstate $\psi$ of the Hamiltonian $\widehat H$ with 
eigenenergy $E$
\begin{equation}
 \widehat H: = \widehat T + \widehat V(r) \,\,,
\label{eqA9}
\end{equation}
%
\begin{equation}
 \widehat H\psi = E\psi
\label{eqA10}
\end{equation}
for the special case that the angular momentum $l$ is the same for all the
states and that two specific eigenenergies $E^{(1)}_\nu$ and $E^{(2)}_\nu$
are very close to the energy $E$ in (\ref{eqA10}).

Then, we may expect that the expansion of $\psi$ in terms of the basis 
states $\varphi^{(1)}_\nu$ and $\varphi^{(2)}_\nu$ may be limited to the two
states with the energies $E^{(1)}_\nu$, $E^{(2)}_\nu$ very close to 
$E_\nu$:
\begin{equation}
 |\psi\rangle = A_\nu|\varphi^{(1)}_\nu \rangle + B_\nu|\varphi^{(2)}_\nu\rangle\,.
\label{eqA11}
\end{equation}
In principle, we would have to sum over all eigenstates $\varphi^{(1)}_\nu$ and 
$\varphi^{(2)}_\nu$ on the righthand side of (\ref{eqA11}), but of all the 
expansion coefficients $A_\nu$, $B_\nu$ only the ones corresponding to the 
energies $E^{(1)}_\nu$, $E^{(2)}_\nu$ closest to $E$ are expected to be large 
and all the others to be much smaller.

Henceforth, we leave away the subscript $\nu$ for simplifying the notation.
It is convenient to introduce the dual basis states $\tilde\varphi^{(1))}$,
$\tilde\varphi^{(2))}$ defined by the orthogonality conditions
\begin{equation}
 \langle\tilde\varphi^{(i)}|\varphi^{(j)}\rangle = \delta_{ij}\,\,,
\label{eqA12} 
\end{equation}
where $i$ and $j$ can be 1 or 2. $\tilde\varphi^{(1)}$ and $\tilde\varphi^{(2)}$ 
can be obtained by diagonalizing the inverse of the overlap matrix 
\begin{equation}
 M^{(ij)} = \langle\varphi^{(i)}|\varphi^{(j)}\rangle \,\,.
\label{eqA13}
\end{equation}
It is easily seen that the Schr\"odinger Eq. (\ref{eqA10}) together with
the assumption (\ref{eqA11}) leads to the two coupled linear equations
\begin{equation}
 \left(\begin{array}{cc}
 (E^{(1)} - E + V^{11}_2) & V^{12}_1 \\
 V^{21}_2 & (E^{(2)} - E + V^{22}_1
\end{array}\right)
\left(\begin{array}{c}
 A \\
 B\end{array}\right) = 0 \,\,,
\label{eqA14}
\end{equation}
where the matrix elements are defined as follows:
\begin{equation}
 V^{ij}_{1,2}: = \langle\tilde\varphi^{(i)}|V_{1,2}|\varphi^{(j)}\rangle \,\,.
\label{eqA15}
\end{equation}
Introducing the abbreviations
\begin{equation}
 \tilde E^{(1)}: = E^{(1)} + V^{11}_2  \,\,,
\label{eqA16}
\end{equation}
%
\begin{equation}
  \tilde E^{(2)}: = E^{(2)} + V^{22}_1
\label{eqA17}
\end{equation}
the solutions $E_{\pm}$ of the eigenvalue equation
\begin{equation}
 (\tilde E^{(1)} - E)(\tilde E^{(2)} - E) - V^{12}_1 \cdot V^{21}_2 = 0
\label{eqA18}
\end{equation}
are given by
\begin{equation}
 E_{\pm} = {\tilde E^{(1)} + \tilde E^{(2}) \over 2} \pm {1\over 2}
 \sqrt{(\tilde E^{(1)} - \tilde E^{(2)})^2 + 4V^{12}_1 V^{21}_2}\,\,.
\label{eqA19}
\end{equation}
Writing the two states $|\psi\rangle$ in the form
\begin{equation}
 |\psi_{\pm}\rangle = A_{\pm} \cdot \left\{|\varphi^{(1)}\rangle \pm 
 \lambda_\pm |\varphi^{(2)}\rangle \right\}\,\,,
\label{eqA20}
\end{equation}
where 
\begin{equation}
 \lambda_{\pm}: = {B_{\pm}\over A_{\pm}}\,\,,
\label{eqA21}
\end{equation}
one obtains $\lambda$ from the Eq. (\ref{eqA14}) and the amplitudes $A_{\pm}$
from the normalisation of $|\psi_{\pm}\rangle$.

The physically interesting quantities are the probabilities $W^{(1,2)}_+$ and 
$W^{(1,2)}_-$ to find the system described by $|\psi_{\pm}\rangle$ in the 
substate $\varphi^{(1)}$ or $\varphi^{(2)}$. These probabilities are given by
the mean-values of the projection operators
\begin{equation}
\widehat P_{(1,2)}: = |\varphi^{(1,2)} \rangle\langle\tilde\varphi^{(1,2)}|\,\,.
\label{eqA23}
\end{equation}
It is easily seen that the probabilities are given by the following equations
\begin{equation}
 W^{(1)}_+ = \langle\psi_+|\widehat P_{(1)}|\psi_+\rangle = 
 {1 + \lambda_+ M^{12}\over 1 + 2\lambda_+ M^{12} + \lambda_+^2}\,\,,
\label{eqA24}
\end{equation}
%
\begin{equation}
 W^{(2)}_+ = \langle\psi_+|\widehat P_{(2)}|\psi_+\rangle = 
 {\lambda_+(\lambda_+ + M^{12}) \over 1 + 2\lambda_+ M^{12} + \lambda_+^2}\,\,,
\label{eqA25}
\end{equation}
%
\begin{equation}
 W^{(1)}_- = \langle\psi_-|\widehat P_{(1)}|\psi_-\rangle = 
 {1 - \lambda_- M^{12} \over 1 - 2\lambda_- M^{12} + \lambda_-^2}\,\,,
\label{eqA26}
\end{equation}
%
\begin{equation}
 W^{(2)}_- = \langle\psi_-|\widehat P_{(2)}|\psi_-\rangle = 
 {\lambda_-(\lambda_- - M^{12}) \over 1 - 2\lambda_- M^{12} + \lambda_-^2}\,\,,
\label{eqA27}
\end{equation}
where $M^{12}$ is the overlap of the two basis states
\begin{equation}
 M^{12} = \langle\varphi^{(1)}|\varphi^{(2)}\rangle
\label{eqA28}
\end{equation}
and the quantities $\lambda_{\pm}$ by
\begin{equation}
 \lambda_{\pm} = {1\over 2V^{12}} \left\{-\Delta E^{(1,2)} \pm 
 \sqrt{(\Delta E^{(1,2)})^2 + 4V^{12}_1 V^{21}_2}\right\}\,\,,
\label{eqA29}
\end{equation}
where
\begin{equation}
 \Delta E: = E^{(1)} - E^{(2)}\,.
\label{eqA30}
\end{equation}
It is seen that for
\begin{equation}
 |\Delta E|^2 \gg 4|V^{12}_1 \cdot V^{21}_2| \,\,,
\label{eqA31}
\end{equation}
i.e. far away from the crossing of the energies $E^{(1)}$ and $E^{(2)}$, one of
the ratios $\lambda_{\pm}$ is much larger than the other.

\begin{figure}[t]
\label{fig12}
\begin{center}
\includegraphics[width=7cm]{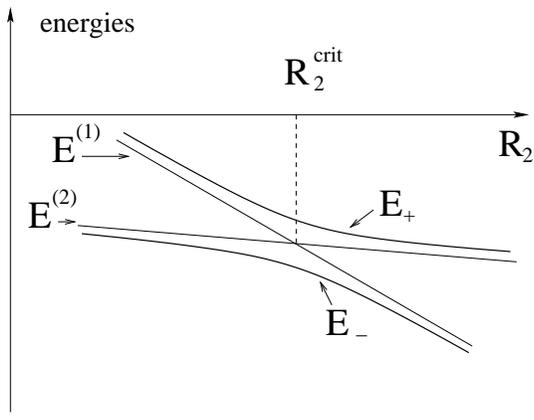}
\caption{Schematic plot of the Landau-Zener crossing of an external and
an internal energy $E^{(1,2)}$ yielding the class 3-state energies $E_{\pm}$.}
\end{center}
\end{figure}

For $\Delta E > 0$ one finds
\begin{equation}
 \lambda_+ \approx {V^{21}_2 \over \Delta E}\,\,,
\label{eqA32}
\end{equation}
%
\begin{equation}
 \lambda_- \approx -\left[{\Delta E\over V^{12}_1} + {V^{21}_2\over \Delta E}\right]
\label{eqA33}
\end{equation}
and for $\Delta E < 0$ the reversed result
\begin{equation}
\lambda_+ \approx  {\Delta E\over V^{12}_1} + {V^{21}_2\over |\Delta E|}\,\,,
\label{eqA34}
\end{equation}
%
\begin{equation}
 \lambda_- \approx -{V^{21}_2 \over |\Delta E|}\,.
\label{eqA35}
\end{equation}
This shows that a predominantly externally localized state turns into a 
predominantly internally localized one as the energies $E^{(1)}$ and $E^{(2)}$ 
cross as a function of $R_2$ or $Z$. The coupled true states $\psi_{\pm}$  are thus
seen to change their nature as a result of the crossing of the energies
$E^{(1)}$, $E^{(2)}$ of a class 1 and a class 2 state. Close to the crossing
point the states $\psi_{\pm}$ are seen to be ''delocalized'', containing 
comparable contributions of the two basis states $\varphi^{(1,2)}$.

This is in essence a Landau-Zener crossing. In Fig.~12 we show a schematic
picture of the energies $E^{(1)}$, $E^{(2)}$, $E_+$, and $E_-$ as a function 
of the the parameter $R_2$ or $Z$.

\end{document}